\documentclass[11pt]{article}
\usepackage{amsmath,amssymb}
\usepackage[dvipdfm]{graphicx}
\usepackage{float}

\makeatletter
\renewcommand{\section}{%
 \@startsection{section}{1}{\z@}{-3.5ex \@plus -1ex \@minus -.2ex}{2.3ex \@plus.2ex}%
  {\normalfont\Large\bfseries}}
\makeatother

\makeatletter
\renewcommand{\subsection}{%
 \@startsection{subsection}{1}{\z@}{-3.5ex \@plus -1ex \@minus -.2ex}{2.3ex \@plus.2ex}%
  {\normalfont\large\bfseries}}
\makeatother

\makeatletter
\def\appendix{
\def\theequation{\Alph{section}\arabic{equation}}
\setcounter{equation}{0}
\def\thesection{\Alph{section}}
\@addtoreset{equation}{section}
\setcounter{section}{0}
\def\@seccntformat##1{
\@nameuse{prefix@##1}
\@nameuse{the##1}
\@nameuse{postfix@##1}\quad}
\def\prefix@section{Appendix~}
} 
\makeatother

\makeatletter
\renewenvironment{thebibliography}[1]
{\section*{\refname\@mkboth{\refname}{\refname}}%
   \list{\@biblabel{\@arabic\c@enumiv}}%
        {\settowidth\labelwidth{\@biblabel{#1}}%
         \leftmargin\labelwidth
         \advance\leftmargin\labelsep
	 \setlength\itemsep{+3pt}%
	 \setlength\baselineskip{11pt}%
         \@openbib@code
         \usecounter{enumiv}%
         \let\p@enumiv\@empty
         \renewcommand\theenumiv{\@arabic\c@enumiv}}%
   \sloppy
   \clubpenalty4000
   \@clubpenalty\clubpenalty
   \widowpenalty4000%
   \sfcode`\.\@m}
  {\def\@noitemerr
    {\@latex@warning{Empty `thebibliography' environment}}%
   \endlist}
\makeatother

\renewcommand{\thefootnote}{\fnsymbol{footnote}}

\begin{document}

\title{{\Large \textbf{Pair creation in boost-invariantly expanding electric fields \\
               and two-particle correlations }} }
\author{{\normalsize Naoto Tanji}\footnote{\textit{E-mail address}: tanji@nt1.c.u-tokyo.ac.jp } }
\date{ \normalsize{\textit{Institute of Physics, University of Tokyo, Komaba, Tokyo 153-8902, Japan} } } 

\maketitle

\renewcommand{\thefootnote}{$*$\arabic{footnote}}

\begin{abstract}
Pair creation of scalar particles in a boost-invariant electric field which is confined in the forward light cone is studied. 
We present the proper-time evolution of momentum distributions of created particles,
which preserve the boost invariance of the background field. 
The two-particle correlation of the created particles is also calculated. 
We find that long-range rapidity correlations may arise from the Schwinger mechanism in the boost-invariant electric field. 
\end{abstract}

\vspace{-280pt}
\begin{flushright}
UT-Komaba/10-8
\end{flushright}
\vspace{240pt}

\section{Introduction} \label{sec:intro}
Boost invariance in a longitudinal beam direction is a key ingredient in theoretical descriptions of 
ultrarelativistic heavy-ion collisions. 
In 1983, Bjorken \cite{Bjorken} introduced a boost-invariant initial condition for Landau's hydrodynamic model \cite{Landau}.
Because the hydrodynamic equations as well respect the boost symmetry, the following space-time evolution of the system
is also boost-invariant, which leads to the central plateau in the particle rapidity distributions. 
Owing to this symmetry,
the fluid evolution is described by the fluid proper time $\tau =\sqrt{t^2 -z^2}$ and the transverse coordinates 
$\mathbf{x}_{\! \perp}$ and does not depend on the space-time rapidity $\eta =\frac{1}{2} \ln \frac{t+z}{t-z}$,
where the longitudinal beam direction is taken along the $z$ axis. 

Before the hydrodynamic evolution, the system of ultrarelativistic heavy-ion collisions is also assumed to 
hold the boost symmetry, 
because shortly after the collision, the nuclei are highly Lorentz-contracted pancakes receding from the collision point
at nearly the speed of light, so that there is no scale in the longitudinal direction. 
For example the flux-tube model (the Low--Nussinov model \cite{Low,Nussinov}), 
which assumes the generation of longitudinal color-electric fields just after high-energy particle collisions, 
comprehends the boost invariance. 
Also in the framework of the color glass condensate, the formation of boost-invariant longitudinal electric fields 
and the longitudinal magnetic fields as well is predicted, of which the state is called glasma \cite{Lappi-Mclerran}. 
Such electric fields would depend on $\tau$ and $\mathbf{x}_{\!\perp}$ but
do not depend on the space-time rapidity $\eta$.  

Starting from the boost-invariant classical fields, how does the system evolve into a boost-invariant and 
locally thermalized quark-gluon plasma? 
One possible approach to this problem is solving the classical Yang--Mills equations and investigating the dynamics
of the classical electromagnetic fields \cite{Lappi-Mclerran,Gatoff-K-V,Fujii-Itakura}. 
Because of the expansion of the system, the electromagnetic fields decay in time. 
To understand the formation process of the quark-gluon plasma, however, the notion of particle or quantized fields is necessary. 
In particular, the classical Yang--Mills equations cannot describe the quark production. 
If one introduces charged quantum fields interacting with the classical electric field into the system evolution,
the system becomes unstable against particle production via the Schwinger mechanism \cite{Schwinger}. 

The Schwinger mechanism, nonperturbative particle pair creation from a classical field,
has been studied as a mechanism of matter formation in heavy-ion collisions \cite{Kajantie-Matsui,Gatoff-K-M,Kluger1991-1993}. 
However, the original Schwinger mechanism treats spatially uniform electric fields. 
In contrast, electric fields expected to be generated in the early stage of heavy-ion collisions exist only
between two nuclei receding from each other at a velocity close to the speed of light. 
In an ideal situation where the two nuclei run at exactly the speed of light,
the electric fields span only inside the forward light cone.  

In the case of spatially homogeneous fields, the semiclassical tunneling description of pair creation
implies that particles and antiparticles are created with vanishing longitudinal momentum. 
This has been assured by field-theoretical calculations \cite{Tanji09}. 
Clearly, these particle-antiparticle pairs break the boost invariance of the background field 
by introducing one specific frame, namely their center of mass frame. 
How is this situation modified if an electric field exists only inside the light cone? 

In this paper, we investigate the Schwinger mechanism under an electric field which localizes in the forward light cone,
focusing on its real-time dynamics and the Lorentz-boost invariance of the system. 
To get the real-time description, we introduce time-dependent particle definitions \cite{Tanji09,Tanji10}. 
The canonical quantization is done on an equal-$\tau$ line instead of an equal-time line 
in order to give a description preserving the boost invariance. 
The proper quantization procedure in the boost-invariant curvilinear coordinates \cite{Sommerfield} provides a boost-invariant description
of the pair creation. 
To concentrate on revealing the boost-invariant dynamics of the particle production, we treat the scalar QED under constant electric fields
which is independent of $\eta$ and neglect the back-reaction of pair creation to the background field. 
We drop the dependence of the electric fields on $\tau$ and $\mathbf{x}_{\!\perp}$ for simplicity. 
Under the constant electric field, we can derive the analytic solution for the distribution of created particles.
Extension to the more general case, quark pair creation under a color-electric field and taking the back-reaction into account,
is possible in a similar way as that in Ref.~\cite{Tanji10}. 

Pair production in electric fields confined to a bounded region has been studied by Martin and Vautherin \cite{Martin-Vautherin}. 
They calculated the pair creation rate by using the Balian-Bloch expansion of Green's functions 
and concluded that if the boundaries move with the speed of light, finite-size correlations vanish and the pair creation rate
is the same as Schwinger's original one.  
In their calculation, information from outside the light cone is not used. 
However, it should be used as an initial condition on the forward light cone
because inside and outside the light cone are causally connected. 
Furthermore, calculating the pair creation rate is not sufficient to reveal the time evolution of the system \cite{Tanji09}. 
To properly describe the system evolution, we must set an initial condition and investigate the time evolution of the quantum fields. 
A study on this line has been carried out by Cooper \textit{et al.} \cite{Cooper1993,Mihaila08}. 
Because the electric field is absent outside the forward light cone, there is no particle created. 
Therefore, the initial state on the light cone should be a vacuum. 
In Ref.~\cite{Cooper1993}, the initial vacuum is defined so that a charge current, which is regularized by the adiabatic regularization,
vanishes at an initial time.  

However, in quantum field theory, there is an ambiguity in identifying a vacuum. 
Requiring the absence of a particle or current at an initial time is not sufficient to determine one initial vacuum.  
Because we work in the flat Minkowski space, the vacuum must hold the Poincar\'{e} invariance, 
even if we use the curvilinear coordinates. 
Hence, the vacuum we should employ for the particle mode defined in the $\tau$-$\eta$ coordinates must be the same as
that associated with the usual quantization of free fields in the Cartesian coordinates. 
In other words, because particles captured by detectors are considered as
the particle modes associated with the usual plane wave expansion of the field operator in the Cartesian coordinates, 
the state before the collision should be the vacuum for that particle mode. 

Because the configuration of the electric field is not symmetric under translation along the beam direction ($z$ axis) 
but symmetric under translation of the space-time rapidity $\eta$, 
particles created from the electric field are eigenstates of the momentum conjugate to $\eta$ 
(which will be denoted as $\lambda$ in this paper), instead of the usual longitudinal momentum $p_z$. 
In Ref.~\cite{Cooper1993}, the relation between these two kinds of momentum variables is given based on 
a coordinate transformation in classical mechanics, 
and the rapidity independence of particle spectra is shown. 
In contrast, we will derive a relation between the two eigenstates of these momenta as an operator relation in a field-theoretical sense
and show that the eigenstate of the momentum conjugate to $\eta$ is a quantum-mechanical superposition of the eigenstate 
of the longitudinal momentum $p_z$. 
This prescription enables us to study multiparticle correlations. 
Because particles are created as a coherent superposition of several momentum modes, 
they are correlated in the momentum (rapidity) space. 
Two-particle correlations between particles produced in the early stage of heavy-ion collisions \cite{Dumitru08}
attract interests in the context of the near-side ridge phenomena observed in heavy-ion collisions at 
the Relativistic Heavy Ion Collider (RHIC)
\cite{STAR05,PHENIX08,PHOBOS10}
and recently in proton-proton collisions at the LHC \cite{CMS}. 
Multiparticle correlations arising from the Schwinger mechanism in spatially homogeneous electric fields have been studied 
by Fukushima, Gelis and Lappi \cite{Fukushima09}. 
We will investigate how that correlation is modified if an electric field spans only inside the forward light cone.   

This paper is organized as follows. 
In Sec.~\ref{sec:quantization}, we review the field quantization in terms of the $\tau$-$\eta$ coordinates.  
Discussions in this section are to a large degree based on Ref.~\cite{Sommerfield}. 
Because of the translational invariance to the $\eta$ direction, particle modes are parametrized 
by the momentum conjugate to $\eta$. 
In Sec.~\ref{sec:meaning}, the physical meaning of the momentum conjugate to $\eta$ is investigated. 
After these provisions, in Sec.~\ref{sec:creation}, 
we will study particle production in the boost-invariant electric field which is confined in the forward light cone. 
The proper-time evolution of the momentum distribution functions is displayed there. 
Finally, two-particle correlations between particles created in the boost-invariant electric field are investigated 
in Sec.~\ref{sec:correlation}. 

\section{Quantum fields in the $\tau$-$\eta$ coordinates} \label{sec:quantization} 
In this section, we will review the canonical quantization in terms of the curvilinear $\tau$-$\eta$ coordinates,  
as a basis to study the pair creation in a boost-invariant electric field which spans only between two charged plates
receding from each other at the speed of light. 

The $\tau$-$\eta$ coordinates $(\tau,\eta,\mathbf{x}_{\! \perp})$ are related with the Cartesian coordinates as
\begin{equation}
\tau = \sqrt{t^2 -z^2} , \ \eta = \frac{1}{2} \ln \left( \frac{t+z}{t-z} \right) , \ \mathbf{x}_{\! \perp} = (x,y) .
\end{equation}
In terms of the $\tau$-$\eta$ coordinates, the Klein--Gordon equation is written as
\begin{equation}
\left( \tau ^2 \frac{\partial ^2}{\partial \tau ^2} +\tau \frac{\partial}{\partial \tau} 
      -\frac{\partial ^2}{\partial \eta ^2} -\tau ^2 \frac{\partial ^2}{\partial x_{\! \perp} ^2} +m^2 \tau ^2 \right) 
      \phi (\tau ,\eta ,\mathbf{x} _{\! \perp}) =0 . \label{KG1}
\end{equation}
$\phi $ is a charged scalar field with a mass $m (\neq 0)$ and a charge $e$.
We quantize the field $\phi$ on an equal $\tau$-line instead of an equal $t$-line, imposing the canonical commutation relation
\begin{equation}
\left[ \phi (\tau ,\eta ,\mathbf{x} _{\! \perp}) ,\pi (\tau ,\eta ^\prime ,\mathbf{x} _{\! \perp} ^\prime ) \right] 
 = i\delta (\eta -\eta ^\prime )\delta ^2 (\mathbf{x} _{\! \perp} -\mathbf{x} _{\! \perp} ^\prime ), \label{comm1}
\end{equation}
where 
$\pi (\tau ,\eta ,\mathbf{x} _{\! \perp}) = \tau \frac{\partial}{\partial \tau} \phi ^\dagger (\tau ,\eta ,\mathbf{x} _{\! \perp}) $ 
is canonical conjugate momentum. 

Because of the translational invariance to the $\mathbf{x} _{\! \perp} $ and $\eta$ directions, 
a solution of Eq.~\eqref{KG1} can be expanded by Fourier modes as 
\begin{equation}
\phi (\tau ,\eta ,\mathbf{x} _{\! \perp}) = \int \! d\lambda d^2 p_{\! \perp} \ \tilde{\phi} _{\mathbf{p} _{\! \perp} ,\lambda } (\tau)
 \frac{1}{\sqrt{(2\pi )^3 } } e^{i\mathbf{p} _{\! \perp} \cdot \mathbf{x} _{\! \perp} } e^{i\lambda \eta} . \label{Fourier}
\end{equation}
This equation tells us that a particle mode defined on an equal $\tau$ line can be parametrized by
quantum numbers $\mathbf{p} _{\! \perp}$ and $\lambda$. 
To define this particle mode, let us decompose the field operator into 
a \lq\lq positive frequency\rq\rq\ part and a \lq\lq negative frequency\rq\rq\ part:
\begin{equation}
\begin{split}
\phi (\tau ,\eta ,\mathbf{x} _{\! \perp}) 
 &= \int \! d\lambda d^2 p_{\! \perp} \left[ 
    \chi ^+ _{\mathbf{p} _{\! \perp} ,\lambda } (\tau ) \mathfrak{a}_{\mathbf{p} _{\! \perp} ,\lambda } 
    +\chi ^- _{\mathbf{p} _{\! \perp} ,\lambda } (\tau ) \mathfrak{b}_{-\mathbf{p} _{\! \perp} ,-\lambda } ^\dagger 
    \right] \frac{1}{\sqrt{(2\pi )^3 } } e^{i\mathbf{p} _{\! \perp} \cdot \mathbf{x} _{\! \perp} } e^{i\lambda \eta} \\
 &= \int \! d\lambda d^2 p_{\! \perp} \left[ 
    \phi ^+ _{\mathbf{p} _{\! \perp} ,\lambda } (\tau ,\eta ,\mathbf{x} _{\! \perp}) 
    \mathfrak{a}_{\mathbf{p} _{\! \perp} ,\lambda } 
    +\phi ^- _{\mathbf{p} _{\! \perp} ,\lambda } (\tau ,\eta ,\mathbf{x} _{\! \perp}) 
    \mathfrak{b}_{-\mathbf{p} _{\! \perp} ,-\lambda } ^\dagger \right] , \label{phi}
\end{split}
\end{equation}
where 
\begin{equation}
\phi ^\pm _{\mathbf{p} _{\! \perp} ,\lambda } (\tau ,\eta ,\mathbf{x} _{\! \perp}) 
 = \chi ^\pm _{\mathbf{p} _{\! \perp} ,\lambda } (\tau ) 
   \frac{1}{\sqrt{(2\pi )^3 } } e^{i\mathbf{p} _{\! \perp} \cdot \mathbf{x} _{\! \perp} } e^{i\lambda \eta} \label{phi-chi}
\end{equation}
are c-number solutions of the field equation \eqref{KG1}. 
The mode functions $\phi ^\pm _{\mathbf{p} _{\! \perp} ,\lambda } (\tau ,\eta ,\mathbf{x} _{\! \perp}) $ 
are set to obey the orthonormal conditions
\begin{gather}
( \phi ^+ _{\mathbf{p} _{\! \perp} ,\lambda } ,\phi ^+ _{\mathbf{p} _{\! \perp} ^\prime ,\lambda ^\prime } )_\tau
 = \delta (\lambda -\lambda ^\prime )\delta ^2 (\mathbf{p} _{\! \perp} -\mathbf{p} _{\! \perp} ^\prime ), \notag \\
( \phi ^- _{\mathbf{p} _{\! \perp} ,\lambda } ,\phi ^- _{\mathbf{p} _{\! \perp} ^\prime ,\lambda ^\prime } )_\tau
 = -\delta (\lambda -\lambda ^\prime )\delta ^2 (\mathbf{p} _{\! \perp} -\mathbf{p} _{\! \perp} ^\prime ), \label{normal1} \\
( \phi ^+ _{\mathbf{p} _{\! \perp} ,\lambda } ,\phi ^- _{\mathbf{p} _{\! \perp} ^\prime ,\lambda ^\prime } )_\tau
 = 0 \ . \notag
\end{gather}
The inner product $( \phi _1 ,\phi _2 )_\tau $ is defined as 
\begin{equation}
( \phi _1 ,\phi _2 )_\tau \equiv i\int _{\tau =\text{const} } \hspace{-25pt} d\eta d^2 x_{\! \perp} \ 
   \tau \left( \phi _1 ^\dagger \overleftrightarrow{\frac{d}{d\tau}} \phi _2 \right) ,
%
\end{equation}
where
\begin{equation}
\phi _1 ^\dagger \overleftrightarrow{\frac{d}{d\tau}} \phi _2
 \equiv \phi _1 ^\dagger \cdot {\frac{d}{d\tau}} \phi _2 -{\frac{d}{d\tau}}\phi _1 ^\dagger \cdot \phi _2 \ .
\end{equation}
Owing to the orthonormal conditions \eqref{normal1}, the commutation relations
\begin{gather}
[\mathfrak{a}_{\mathbf{p} _{\! \perp} ,\lambda } ,\mathfrak{a}_{\mathbf{p} _{\! \perp} ^\prime ,\lambda ^\prime } ^\dagger ] 
 = \delta (\lambda -\lambda ^\prime )\delta ^2 (\mathbf{p} _{\! \perp} -\mathbf{p} _{\! \perp} ^\prime ), \notag \\
[\mathfrak{b}_{\mathbf{p} _{\! \perp} ,\lambda } ,\mathfrak{b}_{\mathbf{p} _{\! \perp} ^\prime ,\lambda ^\prime } ^\dagger ] 
 = \delta (\lambda -\lambda ^\prime )\delta ^2 (\mathbf{p} _{\! \perp} -\mathbf{p} _{\! \perp} ^\prime ), \label{comm2} \\
[\mathfrak{a}_{\mathbf{p} _{\! \perp} ,\lambda } ,\mathfrak{b}_{\mathbf{p} _{\! \perp} ^\prime ,\lambda ^\prime } ] 
 = [\mathfrak{a}_{\mathbf{p} _{\! \perp} ,\lambda } ^\dagger ,\mathfrak{b}_{\mathbf{p} _{\! \perp} ^\prime ,\lambda ^\prime } ^\dagger ] 
 = 0 , \notag 
\end{gather}
are deduced from the canonical commutation relation \eqref{comm1},
and the operators $\mathfrak{a}_{\mathbf{p} _{\! \perp} ,\lambda } ^\dagger$ and 
$\mathfrak{b}_{\mathbf{p} _{\! \perp} ,\lambda }^\dagger$ acquire the role of
the creation operator of a particle with quantum numbers $\mathbf{p} _{\! \perp} $ and $\lambda $.
Furthermore, the expression of the charge operator in terms of 
$\mathfrak{a}_{\mathbf{p} _{\! \perp} ,\lambda }$ and $\mathfrak{b}_{\mathbf{p} _{\! \perp} ,\lambda }$
\begin{equation}
\begin{split}
\hat{Q} &= e(\phi ,\phi )_\tau \\
 &= e\int \! d\lambda d^2 p_{\! \perp} \left[ \mathfrak{a}_{\mathbf{p} _{\! \perp} ,\lambda } ^\dagger 
    \mathfrak{a}_{\mathbf{p} _{\! \perp} ,\lambda }
    -\mathfrak{b}_{\mathbf{p} _{\! \perp} ,\lambda } \mathfrak{b}_{\mathbf{p} _{\! \perp} ,\lambda } ^\dagger \right]
\end{split}
\end{equation}
indicates that $\mathfrak{a}_{\mathbf{p} _{\! \perp} ,\lambda } ^\dagger$ is the creation operator of a particle with charge $+e$
and $\mathfrak{b}_{\mathbf{p} _{\! \perp} ,\lambda } ^\dagger$ is the creation operator of an antiparticle with charge $-e$. 

In order that Eq.~\eqref{phi} is a solution of the Klein--Gordon equation \eqref{KG1}, 
$\chi ^\pm _{\mathbf{p} _{\! \perp} ,\lambda } (\tau ) $ must satisfy
\begin{equation}
\left( \tau ^2 \frac{\partial ^2}{\partial \tau ^2} +\tau \frac{\partial}{\partial \tau} 
      +\lambda ^2 +m_{\! \perp} ^2 \tau ^2 \right) \chi ^\pm _{\mathbf{p} _{\! \perp} ,\lambda } (\tau ) =0 , \label{KG2}
\end{equation}
where $m_{\! \perp}$ is the transverse mass defined by $m_{\! \perp} ^2 = m^2 +p_{\! \perp} ^2$. 
Note that the role of transverse degrees of freedom is only shifting the mass. 
Solutions of Eq.~\eqref{KG2} can be expressed by the Bessel functions. 
For example, we may use the Bessel function of the first kind: 
\begin{equation}
\begin{split}
\chi ^+ _{\mathbf{p} _{\! \perp} ,\lambda } (\tau) &= \sqrt{\frac{\pi}{2\sinh \pi |\lambda |}} J_{-i|\lambda |} (m_{\! \perp} \tau ) \\
\chi ^- _{\mathbf{p} _{\! \perp} ,\lambda } (\tau) &= \sqrt{\frac{\pi}{2\sinh \pi |\lambda |}} J_{i|\lambda |} (m_{\! \perp} \tau ) .
\end{split} \label{Bessel1}
\end{equation}
The normalization factor has been determined by the condition \eqref{normal1} or equivalently
\begin{gather}
i\tau \left( \chi _{\mathbf{p} _{\! \perp} ,\lambda } ^{+\, \dagger} \overleftrightarrow{\frac{d}{d\tau}} 
 \chi _{\mathbf{p} _{\! \perp} ,\lambda } ^+ \right) =1 , \notag \\
i\tau \left( \chi _{\mathbf{p} _{\! \perp} ,\lambda } ^{-\, \dagger} \overleftrightarrow{\frac{d}{d\tau}} 
 \chi _{\mathbf{p} _{\! \perp} ,\lambda } ^- \right) =-1 , \label{normal2} \\
i\tau \left( \chi _{\mathbf{p} _{\! \perp} ,\lambda } ^{+\, \dagger} \overleftrightarrow{\frac{d}{d\tau}} 
 \chi _{\mathbf{p} _{\! \perp} ,\lambda } ^- \right) =0 . \notag 
%
\end{gather}
One can construct a Fock space by using the creation and annihilation operator associated with the solutions \eqref{Bessel1}. 
However, the choice of solutions $\chi _{\mathbf{p} _{\! \perp} ,\lambda } ^\pm $ is not unique. 
There remains the freedom of the Bogoliubov transformation:
\begin{equation}
\begin{split}
\tilde{\chi } _{\mathbf{p} _{\! \perp} ,\lambda } ^+ (\tau ) 
 &= \alpha _{\mathbf{p} _{\! \perp} ,\lambda } \, \chi _{\mathbf{p} _{\! \perp} ,\lambda } ^+ (\tau ) 
   +\beta _{\mathbf{p} _{\! \perp} ,\lambda } ^{*} \, \chi _{\mathbf{p} _{\! \perp} ,\lambda } ^- (\tau ) \\
\tilde{\chi } _{\mathbf{p} _{\! \perp} ,\lambda } ^- (\tau ) 
 &= \alpha _{\mathbf{p} _{\! \perp} ,\lambda } ^{*} \, \chi _{\mathbf{p} _{\! \perp} ,\lambda } ^- (\tau ) 
   +\beta _{\mathbf{p} _{\! \perp} ,\lambda } \, \chi _{\mathbf{p} _{\! \perp} ,\lambda } ^+ (\tau ) ,
\end{split} \label{Bogo1}
\end{equation}
where the coefficients $\alpha _{\mathbf{p} _{\! \perp} ,\lambda } $ and $\beta _{\mathbf{p} _{\! \perp} ,\lambda } $ satisfy
the normalization condition
\begin{equation}
|\alpha _{\mathbf{p} _{\! \perp} ,\lambda } |^2 -|\beta _{\mathbf{p} _{\! \perp} ,\lambda } |^2 = 1 . \label{Bogo_norm}
\end{equation}
For any set of such coefficients, 
$\tilde{\chi } _{\mathbf{p} _{\! \perp} ,\lambda } ^\pm$ are also solutions of Eq.~\eqref{KG2} fulfilling the orthonormal condition
\eqref{normal2}. 
Therefore, there are infinite numbers of ways to decompose the field operator $\phi$ into 
a \lq\lq positive frequency\rq\rq\ part and a \lq\lq negative frequency\rq\rq\ part, 
and each decomposition gives different sets of creation and annihilation operators:
\begin{equation}
\begin{split}
\phi (\tau ,\eta ,\mathbf{x} _{\! \perp})
 &= \int \! d\lambda d^2 p_{\! \perp} \left[ 
    \chi _{\mathbf{p} _{\! \perp} ,\lambda } ^+ (\tau ) \mathfrak{a}_{\mathbf{p} _{\! \perp} ,\lambda } 
    +\chi _{\mathbf{p} _{\! \perp} ,\lambda } ^- (\tau ) \mathfrak{b}_{-\mathbf{p} _{\! \perp} ,-\lambda } ^\dagger 
    \right] \frac{1}{\sqrt{(2\pi )^3 } } e^{i\mathbf{p} _{\! \perp} \cdot \mathbf{x} _{\! \perp} } e^{i\lambda \eta} \\
 &= \int \! d\lambda d^2 p_{\! \perp} \left[ 
    \tilde{\chi} _{\mathbf{p} _{\! \perp} ,\lambda } ^+ (\tau ) \tilde{\mathfrak{a}} _{\mathbf{p} _{\! \perp} ,\lambda } 
    +\tilde{\chi} _{\mathbf{p} _{\! \perp} ,\lambda } ^- (\tau ) \tilde{\mathfrak{b}} _{-\mathbf{p} _{\! \perp} ,-\lambda } ^\dagger 
    \right] \frac{1}{\sqrt{(2\pi )^3 } } e^{i\mathbf{p} _{\! \perp} \cdot \mathbf{x} _{\! \perp} } e^{i\lambda \eta}
\end{split}
\end{equation}
The two kinds of creation and annihilation operators introduced above are related by the following Bogoliubov transformation:
\begin{equation}
\begin{split}
\mathfrak{a}_{\mathbf{p} _{\! \perp} ,\lambda } 
 &= \alpha _{\mathbf{p} _{\! \perp} ,\lambda } \tilde{\mathfrak{a}} _{\mathbf{p} _{\! \perp} ,\lambda }
    +\beta _{\mathbf{p} _{\! \perp} ,\lambda } \tilde{\mathfrak{b}} _{\mathbf{p} _{\! \perp} ,\lambda } ^\dagger \\
\mathfrak{b}_{\mathbf{p} _{\! \perp} ,\lambda } ^\dagger 
 &= \alpha _{\mathbf{p} _{\! \perp} ,\lambda } ^* \tilde{\mathfrak{b}} _{\mathbf{p} _{\! \perp} ,\lambda } ^\dagger
    +\beta _{\mathbf{p} _{\! \perp} ,\lambda } ^* \tilde{\mathfrak{a}} _{\mathbf{p} _{\! \perp} ,\lambda } .
\end{split}
\end{equation}
Because creation and annihilation operators are mixed by the Bogoliubov transformation,
the vacuum $|0\rangle $ defined by $\mathfrak{a}_{\mathbf{p} _{\! \perp} ,\lambda } |0\rangle 
= \mathfrak{b}_{\mathbf{p} _{\! \perp} ,\lambda } |0\rangle = 0$
and the vacuum $|\tilde{0} \rangle $ defined by 
$\tilde{\mathfrak{a}} _{\mathbf{p} _{\! \perp} ,\lambda } |\tilde{0} \rangle 
= \tilde{\mathfrak{b}} _{\mathbf{p} _{\! \perp} ,\lambda } |\tilde{0} \rangle = 0$
are inequivalent:
\begin{equation}
|0\rangle \neq |\tilde{0} \rangle . 
\end{equation}
That is, there are many inequivalent vacua in quantum field theory. 
This problem is not owing to the use of the special coordinates, the $\tau$-$\eta$ coordinates,
but is inherent in quantum field theory. 
In the Cartesian coordinates, however, positive and negative frequency solutions are selected as 
\begin{gather}
\phi _\mathbf{p} ^\pm (t,\mathbf{x} ) = \frac{1}{\sqrt{2\omega _p (2\pi )^3}} e^{\mp i\omega _p t +i\mathbf{p} \cdot \mathbf{x} } 
\label{plane}
\end{gather}
according to the symmetry of the space and time. 
Now, $\omega _p $ is a one-particle energy: $\omega _p = \sqrt{\mathbf{p} ^2 +m^2}$. 
In the $\tau$-$\eta$ coordinates, such a selection of positive and negative frequency solutions is not obvious
because the metric depends on $\tau$ and there is no translational symmetry on $\tau$. 

A proper set of positive and negative frequency solutions, in other words, a proper definition of particle in
the $\tau$-$\eta$ coordinates can be found in the following way \cite{Sommerfield}. 
The use of the $\tau$-$\eta$ coordinates is only a matter of description and should not change the physics. 
Therefore, even if one uses the $\tau$-$\eta$ coordinates, 
a physical vacuum must be the same as that defined in the Cartesian coordinates. 
That is, a \lq\lq positive frequency\rq\rq\ solution in the $\tau $-$\eta $ coordinates must be a superposition
of only the positive frequency solutions defined in the Cartesian coordinates;
the negative frequency cannot be mixed. 
We can construct the solutions satisfying such a condition by using the Hankel functions as follows:
\begin{equation}
 \phi _{\mathbf{p} _{\! \perp} ,\lambda } ^\pm (\tau ,\eta ,\mathbf{x} _{\! \perp}) 
  = \chi _{\mathbf{p} _{\! \perp} ,\lambda } ^\pm (\tau ) 
    \frac{e^{i\mathbf{p} _{\! \perp} \cdot \mathbf{x} _{\! \perp}+i\lambda \eta } }{\sqrt{(2\pi )^3 }} , \label{+-phi0}
\end{equation}
where
\begin{gather}
\chi _{\mathbf{p} _{\! \perp} ,\lambda } ^+ (\tau ) 
 = \frac{\sqrt{\pi }}{2i} e^{\frac{\pi}{2} \lambda } H_{i\lambda } ^{(2)} (m_{\! \perp} \tau ) \label{+chi} \\
\chi _{\mathbf{p} _{\! \perp} ,\lambda } ^- (\tau ) 
 = \left[ \chi _{\mathbf{p} _{\! \perp} ,\lambda } ^+ (\tau ) \right] ^* 
 = -\frac{\sqrt{\pi }}{2i} e^{\frac{\pi}{2} \lambda } H_{-i\lambda } ^{(1)} (m_{\! \perp} \tau ) . \label{-chi}
\end{gather}
Using the integral representation of the Hankel function \cite{Abramowitz} 
\begin{equation}
H_\nu ^{(2)} (z) = \frac{ie^{\pi i\nu /2}}{\pi } \int _{-\infty} ^\infty \! dt \ e^{-iz\cosh t +\nu t} 
 \hspace{10pt} \left[ |\text{Re} \, \nu |<1, \text{Im} \, z <0 \right] , \label{H_int2}
\end{equation} 
one can rewrite 
$\phi _{\mathbf{p} _{\! \perp} ,\lambda } ^\pm (\tau ,\eta ,\mathbf{x} _{\! \perp}) $ 
as follows:
\begin{gather}
\begin{split}
\phi _{\mathbf{p} _{\! \perp} ,\lambda } ^+ (\tau ,\eta ,\mathbf{x} _{\! \perp}) 
 &= \frac{1}{\sqrt{2\pi } } \int _{-\infty } ^\infty \! \frac{dp_z }{\sqrt{\omega _p } } 
    \phi _\mathbf{p} ^+ (t,\mathbf{x} ) e^{i\lambda y_p } 
\end{split} \\
\begin{split}
\phi _{\mathbf{p} _{\! \perp} ,\lambda } ^- (\tau ,\eta ,\mathbf{x} _{\! \perp}) 
 &= \frac{1}{\sqrt{2\pi } } \int _{-\infty } ^\infty \! \frac{dp_z }{\sqrt{\omega _p } } 
    \phi _\mathbf{p} ^- (t,\mathbf{x} ) e^{-i\lambda y_p } , 
\end{split}
\end{gather}
where $y_p$ denotes rapidity corresponding to momentum $\mathbf{p} $: $y_p = \tanh ^{-1} (p_z /\omega _p )$. 
The condition $\text{Im} \, z <0$ in \eqref{H_int2} is satisfied if we suppose that the mass has a small imaginary part
as $m-i\epsilon$, which is a usual prescription in quantum field theory. 
An important point is that the positive frequency and the negative frequency solutions are not mixed with each other. 
In the forward light cone, one can decompose the field operator in two ways by using 
the mode solutions in the Cartesian coordinates
$\phi _\mathbf{p} ^\pm (t,\mathbf{x} ) $ [Eq.~\eqref{plane}]
and in the $\tau$-$\eta$ coordinates
$\phi _{\mathbf{p} _{\! \perp} ,\lambda } ^\pm (\tau ,\eta ,\mathbf{x} _{\! \perp}) $ [Eq.~\eqref{+-phi0}]:
\begin{align}
\phi (t,\mathbf{x} ) &= \int \! d^3 p \left[ \phi _\mathbf{p} ^+ (t,\mathbf{x} ) a_\mathbf{p}
     +\phi _\mathbf{p} ^- (t ,\mathbf{x} ) b_{-\mathbf{p} } ^\dagger \right] \label{expansion01} \\
 &= \int \! d^2 p_{\! \perp} d\lambda \left[ 
    \phi _{\mathbf{p} _{\! \perp} ,\lambda } ^+ (\tau ,\eta ,\mathbf{x} _{\! \perp}) 
    \mathfrak{a}_{\mathbf{p} _{\! \perp} ,\lambda } 
    +\phi _{\mathbf{p} _{\! \perp} ,\lambda } ^- (\tau ,\eta ,\mathbf{x} _{\! \perp}) 
    \mathfrak{b}_{-\mathbf{p} _{\! \perp} ,-\lambda } ^\dagger \right] . \label{expansion02}
\end{align}
Then, the relations between the creation or annihilation operator associated with $\phi _\mathbf{p} ^\pm (t,\mathbf{x} ) $
and that with $\phi _{\mathbf{p} _{\! \perp} ,\lambda } ^\pm (\tau ,\eta ,\mathbf{x} _{\! \perp}) $ are obtained as
\begin{gather}
\begin{split}
a_\mathbf{p} &= (\phi _\mathbf{p} ^+ ,\phi )_t 
 = \frac{1}{\sqrt{2\pi \omega _p }} \int _{-\infty} ^\infty \! d\lambda \ e^{i\lambda y_p } 
   \mathfrak{a}_{\mathbf{p} _{\! \perp} ,\lambda } \ , \\
b_\mathbf{p} ^\dagger &= -(\phi _{-\mathbf{p} } ^- ,\phi )_t 
 = \frac{1}{\sqrt{2\pi \omega _p }} \int _{-\infty} ^\infty \! d\lambda \ e^{i\lambda y_p } 
   \mathfrak{b}_{\mathbf{p} _{\! \perp} ,\lambda } ^\dagger \ .
\end{split} \label{relation}
\end{gather}
Their inverse relations are
\begin{gather}
\begin{split}
\mathfrak{a}_{\mathbf{p} _{\! \perp} ,\lambda } 
 &= \frac{1}{\sqrt{2\pi } } \int _{-\infty} ^\infty \! \frac{dp_z }{\sqrt{\omega _p } } e^{-i\lambda y_p } a_\mathbf{p} \ , \\
\mathfrak{b}_{\mathbf{p} _{\! \perp} ,\lambda } ^\dagger
 &= \frac{1}{\sqrt{2\pi } } \int _{-\infty} ^\infty \! \frac{dp_z }{\sqrt{\omega _p } } e^{-i\lambda y_p } b_\mathbf{p} ^\dagger \ .
\end{split}
\end{gather}
From these equations, we see that the particle mode respecting the boost invariance is a superposition of various momentum modes.
Because the particle modes and the antiparticle modes are not mixed with each other, 
a vacuum defined in the Cartesian coordinates $a_\mathbf{p} |0\rangle =b_\mathbf{p} |0\rangle =0$ is also a vacuum
for the particle mode defined in the $\tau$-$\eta$ coordinates:
\begin{equation}
\mathfrak{a}_{\mathbf{p} _{\! \perp} ,\lambda } |0\rangle = \mathfrak{b}_{\mathbf{p} _{\! \perp} ,\lambda } |0\rangle = 0. 
\end{equation} 
Using these particle modes defined by Eq.~\eqref{relation}, we can properly calculate field-theoretical quantities 
in terms of the $\tau$-$\eta$ coordinates. 

\section{The meaning of the momentum conjugate to the space-time rapidity} \label{sec:meaning}

In the previous section, we have formulated the field quantization in terms of the $\tau$-$\eta$ coordinates 
and derived the relation between the particle modes respecting the boost invariance and those associated with 
the usual plane wave [Eq.~\eqref{relation}]. 
In that process, the quantum number $\lambda$ has been introduced as a conjugate variable to $\eta$. 
However, the physical meaning of $\lambda$ is not obvious from the discussions in the previous section. 
We will investigate it in this section. 

\subsection{Classical mechanics} \label{subsec:cm}
First, we study the meaning of $\lambda$ in the framework of the classical mechanics in the $\tau$-$\eta$ coordinates. 
For simplicity, we analyze a (1+1)-dimensional system. 
The action of a free particle is
\begin{equation}
S_0 = -m\int \! dt \sqrt{1-\dot{z} ^2 } = -m\int \! d\tau \sqrt{1-\tau ^2 \dot{\eta} ^2 } ,
\end{equation}
where
\begin{align}
\dot{z} &\equiv \frac{dz}{dt} \\
\dot{\eta} &\equiv \frac{d\eta }{d\tau } .
\end{align} 
Hence, the Lagrangian in terms of the $\tau$-$\eta$ coordinates is
\begin{equation}
L_\tau = -m\sqrt{1-\tau ^2 \dot{\eta} ^2 } .
\end{equation}
Motion of the particle is expressed as $z=z(t)$ or $\eta =\eta (\tau )$, both of which give the same motion 
if the particle is in the forward light cone.   
The momentum conjugate to $\eta $ is
\begin{equation}
\lambda = \frac{\partial L_\tau}{\partial \dot{\eta} } 
        = \frac{m\tau ^2 \dot{\eta} }{\sqrt{1-\tau ^2 \dot{\eta} ^2 }} . \label{peta1}
\end{equation}
Using the relation
\begin{equation}
\dot{\eta} = \frac{1}{\tau} \frac{t\dot{z} -z}{t-z\dot{z} } ,
\end{equation}
one can rewrite Eq.~\eqref{peta1} as
\begin{equation}
\lambda = \frac{m(\dot{z}t-z)}{\sqrt{1-\dot{z} ^2}}
       = p_z t -\omega _p z ,
\end{equation}
where $p_z $ is the momentum conjugate to $z$ and $\omega _p $ is the energy of the particle:
\begin{align}
p_z &= \frac{m\dot{z}}{\sqrt{1-\dot{z} ^2}}  \\
\omega _p &= \frac{m}{\sqrt{1-\dot{z} ^2}} \ .
\end{align}

To make the meaning of $\lambda$ clearer, let us boost this system by the velocity $v_z =z(t)/t $ in the $z$ direction,
where $z(t)$ is the position of the particle at time $t$. 
By this boost, $t$ and $z$ are transformed as
\begin{align}
t &\rightarrow t^\prime = \tau \\
z &\rightarrow z^\prime = 0 ,
\end{align}
where the prime denotes a boosted quantity. 
Because $\lambda$ is a boost-invariant quantity,
\begin{equation}
\lambda = p_z t -\omega _p z = p_z ^\prime t^\prime -\omega _p ^\prime z^\prime =  p_z ^\prime \tau \ .
\end{equation}
This equation clearly tells us the meaning of $\lambda$: 
$\lambda /\tau $ is the longitudinal momentum of the particle in the frame which moves with the longitudinal velocity $v_z =z(t)/t $. 
Note that the velocity $v_z =z(t)/t $ is time-dependent unless the motion of the particle is a straight line from the origin 
$z(t) \propto t $. 
That is why $p_z ^\prime =\lambda /\tau $ depends on $\tau $ even if the particle is free and $\lambda $ is constant. 

Under the gauge field $A_\tau=0, A_\eta = \frac{1}{2} E\tau ^2$ giving a constant electric field,
the action of a charged particle is 
\begin{equation}
S = -m\int \! d\tau \sqrt{1-\tau ^2 \dot{\eta} ^2 } -e\int \! d\tau \ A_\eta \dot{\eta} \ ,
\end{equation}
which leads the equation of motion 
\begin{equation}
\frac{d}{d\tau} \left( \lambda -\frac{1}{2} eE\tau ^2 \right) =0 .
\end{equation}
If $\lambda =0$ at $\tau =0$, this equation implies that $p_z ^\prime =\lambda/\tau = \frac{1}{2} eE\tau$. 
It seems that the acceleration the particle feels is half compared with the description in terms of the Cartesian coordinates:
$p_z =eEt$. 
This is because $v_z =z(t)/t $ varies under the electric field
and thereby the frame where the longitudinal momentum $p_z ^\prime =\lambda/\tau $ is defined is also changing in time. 

\subsection{Quantum field theory} \label{subsec:qft}

In this subsection, we investigate the physical meaning of $\lambda$ 
in the frame work of quantum field theory. 
The key of this analysis is Eq.~\eqref{relation}, which relate the particle modes in terms of the $\tau$-$\eta$ coordinates and
those in the Cartesian coordinates. 

Let us define a one-particle state with quantum number $\lambda$ and $\mathbf{q} _{\! \perp} $ as
\begin{equation}
|\mathbf{q} _{\! \perp} ,\lambda \rangle 
 = \sqrt{\frac{\tau (2\pi)^3}{V_\eta}} \mathfrak{a}_{\mathbf{q} _{\! \perp} ,\lambda } ^\dagger |0\rangle ,
\end{equation}
where the normalization factor is introduced so that 
$\langle \mathbf{q} _{\! \perp} ,\lambda |\mathbf{q} _{\! \perp} ,\lambda \rangle =1$, and
$V_\eta $ is a space volume on the $\tau$-constant hypersurface, introduced as
\begin{equation}
\begin{split}
V_\eta &= L_\eta L^2 
  = \int \! \tau d\eta \ \int \! d^2 x_{\! \perp} \\
 &= \tau \int \! d\eta \ e^{i(\lambda -\lambda )\eta} 
    \int \! d^2 x_{\! \perp} \ e^{i(\mathbf{p} _{\! \perp} -\mathbf{p} _{\! \perp} )\cdot \mathbf{x} _{\! \perp} } \\
 &= \tau (2\pi ) ^3 \delta (\lambda -\lambda )\delta ^2 (\mathbf{p} _{\! \perp} -\mathbf{p} _{\! \perp} ) .
\end{split}
\end{equation}
By calculating expectation values of several quantities with this one-particle state, 
we shall examine the meaning of $\lambda $.  

First, we study the expectations of particle number operators. 
The expectation of the number operator in the $\mathfrak{a}_{\mathbf{p} _{\! \perp} ,\lambda } $ basis is
\begin{equation}
\begin{split}
\frac{dN}{d\lambda ^\prime d^2 p_{\! \perp} } 
 &= \langle \mathbf{q} _{\! \perp} ,\lambda |\mathfrak{a}_{\mathbf{p} _{\! \perp} ,\lambda^\prime } ^\dagger
    \mathfrak{a}_{\mathbf{p} _{\! \perp} ,\lambda ^\prime } |\mathbf{q} _{\! \perp} ,\lambda \rangle \\
 &= \frac{\tau (2\pi)^3}{V_\eta} 
    \left[ \delta (\lambda -\lambda ^\prime ) \delta ^2 (\mathbf{p} _{\! \perp} -\mathbf{q} _{\! \perp} ) \right] ^2 \\
 &= \delta (\lambda -\lambda ^\prime ) \delta ^2 (\mathbf{p} _{\! \perp} -\mathbf{q} _{\! \perp} ) .
\end{split}
\end{equation}
This is a trivial result since the expectation is taken by the eigenstate of 
$\mathbf{p} _{\! \perp} $ and $\lambda $,
and does not tell us anything about the physical meaning of the quantum number $\lambda $.  

Next, we study the expectation of the number operator in the $a_\mathbf{q} $-basis: 
\begin{equation}
\begin{split}
\frac{dN}{d^3 p}
 &= \langle \mathbf{q} _{\! \perp} ,\lambda |a_{\mathbf{p} } ^\dagger 
    a_{\mathbf{p} } |\mathbf{q} _{\! \perp} ,\lambda \rangle \\
 &= \frac{1}{2\pi \omega _p } \int \! d\lambda ^\prime e^{-i\lambda ^\prime y_p }
    \int \! d\lambda ^{\prime \prime} e^{i\lambda ^{\prime \prime} y_p } 
    \langle \mathbf{q} _{\! \perp} ,\lambda |\mathfrak{a}_{\mathbf{p} _{\! \perp} ,\lambda ^\prime } ^\dagger 
    \mathfrak{a}_{\mathbf{p} _{\! \perp} ,\lambda ^{\prime \prime} } |\mathbf{q} _{\! \perp} ,\lambda \rangle \\
 &= \frac{1}{\omega _p } \frac{\tau (2\pi)^2}{V_\eta} 
    \left[ \delta ^2 (\mathbf{p} _{\! \perp} -\mathbf{q} _{\! \perp} ) \right] ^2 \\
 &= \frac{1}{\omega _p } \frac{\tau}{L_\eta} \delta ^2 (\mathbf{p} _{\! \perp} -\mathbf{q} _{\! \perp} ) .
\end{split}
\end{equation}
Converting the longitudinal momentum $p_z$ to the rapidity $y_p$, we can get the rapidity distribution
\begin{equation}
\begin{split}
\frac{dN}{dy_p d^2 p_{\! \perp} } &= \omega _p \frac{dN}{d^3 p} \\
 &= \frac{\tau}{L_\eta} \delta ^2 (\mathbf{p} _{\! \perp} -\mathbf{q} _{\! \perp} ) . \label{dNdyone}
\end{split}
\end{equation}
This is independent of rapidity $y_p$. 
The one-particle state $|\mathbf{q} _{\! \perp} ,\lambda \rangle $ contains all rapidity modes 
with equal weight. 
This result is consistent with the boost invariance of
the state $|\mathbf{q} _{\! \perp} ,\lambda \rangle $ in the longitudinal direction. 

Furthermore, because several momentum modes are condensed in the state  $|\mathbf{q} _{\! \perp} ,\lambda \rangle $, 
\begin{equation}
 \langle \mathbf{q} _{\! \perp} ,\lambda |a_{\mathbf{p} } ^\dagger a_{\mathbf{p} ^\prime } |\mathbf{q} _{\! \perp} ,\lambda \rangle 
  = \frac{1}{\sqrt{\omega _p \omega _{p^\prime  } } } e^{-i\lambda (y_p -y_{p^\prime } )} 
    \frac{\tau (2\pi)^2}{V_\eta} 
    \delta ^2 (\mathbf{p} _{\! \perp} -\mathbf{q} _{\! \perp} ) \delta ^2 (\mathbf{p} _{\! \perp} ^\prime -\mathbf{q} _{\! \perp} ) 
\end{equation}
is nonzero even if $p_z \neq p_z ^\prime $. 
This implies that nontrivial correlations in the longitudinal momentum space may arise from the particle production
in the boost-invariant electric field. We will investigate it in Sec.~\ref{sec:correlation}. 

Because the rapidity distribution \eqref{dNdyone} is also independent of $\lambda$, 
the meaning of $\lambda$ is not obtained from it. 
To further investigate the meaning of $\lambda$, we study the energy-momentum tensor.
In the Cartesian coordinates, the energy-momentum tensor for free complex scalar fields is
\begin{equation}
T^{\mu \nu } 
 = \sqrt{-g}
    \left[ \partial ^\mu \phi ^\dagger \partial ^\nu \phi +\partial ^\nu \phi ^\dagger \partial ^\mu \phi 
    -g^{\mu \nu} \left( \partial _\rho \phi ^\dagger \partial ^\rho \phi -m^2 \phi ^\dagger \phi \right) \right] 
\end{equation} 
$(\mu ,\nu =0,1,2,3)$, 
where $g^{\mu \nu} = \text{diag} (1,-1,-1,-1)$ is the metric of the flat Cartesian coordinates system 
and $\sqrt{-g}=\sqrt{|\text{det} g_{\mu \nu} |}=1$. 
Also in the $\tau$-$\eta$ coordinates, it has the same form:
\begin{equation}
 T^{\alpha \beta } 
 = \sqrt{-\mathfrak{g}} 
    \left[ \partial ^\alpha \phi ^\dagger \partial ^\beta \phi +\partial ^\beta \phi ^\dagger \partial ^\alpha \phi 
    -\mathfrak{g}^{\alpha \beta} 
    \left( \partial _\gamma \phi ^\dagger \partial ^\gamma \phi -m^2 \phi ^\dagger \phi \right) \right] 
\end{equation} 
$(\alpha ,\beta =\tau ,1,2,\eta)$, 
where 
$\mathfrak{g}^{\alpha \beta} = \text{diag} (1,-1,-1,-1/\tau^2)$ is the metric of the $\tau$-$\eta$ coordinates system,
and $\sqrt{-\mathfrak{g}} = \sqrt{|\text{det} \mathfrak{g}_{\alpha \beta} |} =\tau $.  

The relations between the differentials in the Cartesian coordinates and those in the $\tau$-$\eta$ coordinates are
\begin{equation}
 \begin{pmatrix}
  \partial ^\tau \\ \partial ^1 \\ \partial ^2 \\ \tau \partial ^\eta
 \end{pmatrix}
 =
 \begin{pmatrix}
  \cosh \eta & 0 & 0 &  -\sinh \eta \\
  0 & 1 & 0 & 0 \\
  0 & 0 & 1 & 0 \\
  -\sinh \eta & 0 & 0 & \cosh \eta
 \end{pmatrix}
 \begin{pmatrix}
  \partial ^0 \\ \partial ^1 \\ \partial ^2 \\ \partial ^3
 \end{pmatrix} . \label{diff}
\end{equation}
Note that the matrix in the right hand side accords with the matrix representing the Lorentz boost along the $z$ axis
with the velocity $v_z =z/t=\tanh \eta$ :
\begin{equation}
 \Lambda ^\mu _{\ \nu} =
 \begin{pmatrix}
  \cosh \eta & 0 & 0 &  -\sinh \eta \\
  0 & 1 & 0 & 0 \\
  0 & 0 & 1 & 0 \\
  -\sinh \eta & 0 & 0 & \cosh \eta
 \end{pmatrix} . \label{boost}
\end{equation}
Therefore, 
\begin{equation}
 \tilde{T} ^{\mu \nu} \equiv \frac{1}{\sqrt{-\mathfrak{g}}} 
 \begin{pmatrix}
  T^{\tau \tau } & T^{\tau 1} & T^{\tau 2} &  \tau T^{\tau \eta} \\
  T^{1 \tau } & T^{1 1} & T^{1 2} &  \tau T^{1 \eta} \\
  T^{2 \tau } & T^{2 1} & T^{2 2} &  \tau T^{2 \eta} \\
  \tau T^{\eta \tau } & \tau T^{\eta 1} & \tau T^{\eta 2} &  \tau ^2 T^{\eta \eta} 
 \end{pmatrix}
\end{equation}
equals the energy-momentum tensor transformed by the Lorentz boost \eqref{boost}:
\begin{equation}
 \tilde{T} ^{\mu \nu} = \Lambda ^\mu _{\ \sigma} \Lambda ^\nu _{\ \rho} T^{\sigma \rho } .
\end{equation} 
For later reference, let us define the $\eta$-frame as the frame which is boosted by \eqref{boost}. 
Then, $\tilde{T} ^{\mu \nu} (\tau,\eta)$ is the energy-momentum tensor observed in the $\eta$-frame. 

The expectation value of $\tilde{T} ^{03} (\tau,\eta)=T^{\tau \eta} (\tau ,\eta)$, 
which represents the momentum density in the $z$ direction observed in the $\eta$-frame, 
with the one-particle state $|\mathbf{p} _{\! \perp} ,\lambda \rangle $ is
\begin{equation}
 \langle \mathbf{p} _{\! \perp} ,\lambda | T^{\tau \eta} (\tau ,\eta) |\mathbf{p} _{\! \perp} ,\lambda \rangle 
  = \frac{\lambda}{\tau} \frac{1}{V_\eta} . \label{T^te}
\end{equation}
This means that the momentum the state $|\mathbf{p} _{\! \perp} ,\lambda \rangle $ contains is $\lambda /\tau$. 
This result is consistent with the result of the classical mechanics in the previous subsection. 
In the present case, however, $\eta$ is a parameter indicating a space point,
while in the classical mechanics $\eta$ is a mechanical variable representing a point where a particle exists. 
Therefore, Eq.~\eqref{T^te} means that momentum $\lambda /\tau$ is distributed at any space point of $\eta$. 

\section{Pair creation in the forward light cone} \label{sec:creation}
In this section, we study the pair creation in a constant electric field which exists only inside the forward light cone. 

The metric of the $\tau$-$\eta$ coordinates has a singularity in $\tau =0$. 
To avoid this singularity, 
we suppose that the electric field is zero when $0 \leq \tau <\tau _0 $ and is switched on at $\tau =\tau _0 >0 $. 
Such an electric field is given by the gauge
\begin{equation}
A_\eta (\tau ) = \begin{cases}
\frac{1}{2} E\tau_0 ^2 & (\tau <\tau _0 ) \\
\frac{1}{2} E\tau ^2 & (\tau \geq \tau _0 )
\end{cases} \label{gauge}
\end{equation} 
and $A_\tau =0$. 
The pure gauge $A_\eta = \frac{1}{2} E\tau_0 ^2 $ at $\tau <\tau _0$ is introduced so that the gauge field is continuous. 
It will be confirmed later that if $\tau_0 $ is sufficiently small,
results of the calculations are insensitive to the values of $\tau_0$.  

The Klein--Gordon equation under this gauge field is
\begin{equation}
\left[ \tau ^2 \partial _\tau ^2 +\tau \partial _\tau -D_\eta ^2 
      -\tau ^2 \partial _{\! \perp} ^2 +m^2 \tau ^2 \right] 
      \phi (\tau ,\eta ,\mathbf{x} _{\! \perp}) =0 , 
\label{KGinE1}
\end{equation} 
where $D_\eta = \partial_\eta +ieA_\eta$. 
A solution of this equation can be expanded in the same way as the free field case:
\begin{equation}
\phi (\tau ,\eta ,\mathbf{x} _{\! \perp} )
 = \int \! d^2 p_{\! \perp} d\lambda \left[ 
    \phi _{\mathbf{p} _{\! \perp} ,\lambda } ^{+\, \text{in}} (\tau ,\eta ,\mathbf{x} _{\! \perp} ) 
    \mathfrak{a}_{\mathbf{p} _{\! \perp} ,\lambda } ^\text{in} 
    +\phi _{\mathbf{p} _{\! \perp} ,\lambda } ^{-\, \text{in}} (\tau ,\eta ,\mathbf{x} _{\! \perp} )
     \mathfrak{b}_{-\mathbf{p} _{\! \perp} ,-\lambda } ^{\text{in} \ \dagger} \right] . \label{expansion1}
\end{equation} 
$\phi _{\mathbf{p} _{\! \perp} ,\lambda } ^{\pm\, \text{in}} (\tau ,\eta ,\mathbf{x} _{\! \perp} ) $ 
is a positive or negative frequency mode function satisfying Eq.~\eqref{KGinE1}. 
The superscripts \lq in\rq\ specify the initial condition for the field:  
At $\tau <\tau _0 $, there is no electric field so that the mode functions are free ones. 
The free solutions at $\tau <\tau_0$ are given by the gauge transformation
$(A_\eta =0\to E\tau_0^2 /2)$ of Eq.~\eqref{+-phi0}: 
\begin{equation}
\begin{split}
\phi _{\mathbf{p} _{\! \perp} ,\lambda } ^{+\, \text{in}} (\tau ,\eta ,\mathbf{x} _{\! \perp}) 
 &= \frac{\sqrt{\pi }}{2i} e^{\frac{\pi}{2} \lambda } H_{i\lambda } ^{(2)} (m_{\! \perp} \tau ) 
   \frac{1}{\sqrt{(2\pi)^3}} e^{i\mathbf{p}_{\! \perp} \cdot \mathbf{x}_{\! \perp} 
   +i(\lambda -\frac{1}{2} E\tau_0^2 )\eta} , \\
\phi _{\mathbf{p} _{\! \perp} ,\lambda } ^{-\, \text{in}} (\tau ,\eta ,\mathbf{x} _{\! \perp}) 
 &= -\frac{\sqrt{\pi }}{2i} e^{\frac{\pi}{2} \lambda } H_{-i\lambda } ^{(1)} (m_{\! \perp} \tau ) 
   \frac{1}{\sqrt{(2\pi)^3}} e^{i\mathbf{p}_{\! \perp} \cdot \mathbf{x}_{\! \perp} 
   +i(\lambda -\frac{1}{2} E\tau_0^2 )\eta} .
\end{split} 
\label{inin}
\end{equation}
The mode functions $\phi _{\mathbf{p} _{\! \perp} ,\lambda } ^{\pm \, \text{in}} $ at $\tau \geq \tau _0 $
are constructed so that $\phi _{\mathbf{p} _{\! \perp} ,\lambda } ^{\pm \, \text{in}} $
and their derivative with respect to $\tau $ are continuous, respectively, at $\tau =\tau _0 $. 
Their explicit forms can be found in the appendix \ref{sec:solutions}. 
By expanding the field operator with these mode functions 
$\phi _{\mathbf{p} _{\! \perp} ,\lambda } ^{\pm \, \text{in}}$ as Eq.~\eqref{expansion1}, 
we can obtain the annihilation operators 
$\mathfrak{a}_{\mathbf{p} _{\! \perp} ,\lambda } ^\text{in}$ and 
$\mathfrak{b}_{\mathbf{p} _{\! \perp} ,\lambda } ^\text{in}$ 
and the associated vacuum state $|0,\text{in} \rangle $, which is defined as
\begin{equation}
\mathfrak{a}_{\mathbf{p} _{\! \perp} ,\lambda } ^\text{in} |0,\text{in} \rangle 
 = \mathfrak{b}_{\mathbf{p} _{\! \perp} ,\lambda } ^\text{in} |0,\text{in} \rangle = 0 .
\end{equation}
We choose this vacuum as the state of this system. 
This choice is a manifestation of the initial condition that there is no particle before the electric field is turned on. 
An important point is that this vacuum is also the vacuum for particles defined in the Cartesian coordinates
as discussed in Sec.~\ref{sec:quantization}. 
Therefore, although the $\tau$-$\eta$ coordinates can cover only inside the forward light cone, 
the absence of particles outside the forward light cone is guaranteed. 

Because we now use the Heisenberg representation, 
the state $|0,\text{in} \rangle $ does not change during time evolution. 
What evolves in time is a definition of particles or, in other words, creation and annihilation operators of particles. 
Such a time-dependent definition of particles is introduced by decomposing the field operator into positive and 
negative frequency instantaneously \cite{Tanji09}: 
\begin{equation} \label{expansion2}
\phi (\tau ,\eta ,\mathbf{x} _{\! \perp} )
 = \int \! d^2 p_{\! \perp} d\lambda \left[ 
   \phi _{\mathbf{p} _{\! \perp} ,\lambda } ^{+\, (\tau _1)} (\tau ,\eta ,\mathbf{x}_{\! \perp}) 
   \mathfrak{a}_{\mathbf{p} _{\! \perp} ,\lambda } (\tau _1) 
   +\phi _{\mathbf{p} _{\! \perp} ,\lambda } ^{-\, (\tau _1)} (\tau ,\eta ,\mathbf{x}_{\! \perp}) 
     \mathfrak{b}_{-\mathbf{p} _{\! \perp} ,-\lambda } ^\dagger (\tau _1) \right] , 
\end{equation}
where $\phi _{\mathbf{p} _{\! \perp} ,\lambda } ^{\pm \, (\tau _1)} (\tau ,\eta ,\mathbf{x}_{\! \perp}) $ 
is a positive or negative frequency solution of the equation of motion \eqref{KGinE1} under the pure gauge 
$A_\eta =A_\eta (\tau =\tau _1 )$: 
\begin{equation}
\begin{split}
\phi _{\mathbf{p} _{\! \perp} ,\lambda } ^{+\, (\tau _1)} (\tau ,\eta ,\mathbf{x}_{\! \perp}) 
 &= \frac{\sqrt{\pi }}{2i} e^{\frac{\pi}{2} \lambda } H_{i\lambda } ^{(2)} (m_{\! \perp} \tau ) 
   \frac{1}{\sqrt{(2\pi )^3 }} e^{i\mathbf{p} _{\! \perp} \cdot \mathbf{x} _{\! \perp}}
   e^{i\left[ \lambda -eA_\eta (\tau _1) \right] \eta } , \\
\phi _{\mathbf{p} _{\! \perp} ,\lambda } ^{-\, (\tau _1)} (\tau ,\eta ,\mathbf{x}_{\! \perp})  
 &= -\frac{\sqrt{\pi }}{2i} e^{\frac{\pi}{2} \lambda } H_{-i\lambda } ^{(1)} (m_{\! \perp} \tau )
   \frac{1}{\sqrt{(2\pi )^3 }} e^{i\mathbf{p} _{\! \perp} \cdot \mathbf{x} _{\! \perp}}
   e^{i\left[ \lambda -eA_\eta (\tau _1) \right] \eta } . 
\end{split} \label{phi^t1}
\end{equation}
The operators $\mathfrak{a}_{\mathbf{p} _{\! \perp} ,\lambda } (\tau )$ and 
$\mathfrak{b}_{\mathbf{p} _{\! \perp} ,\lambda } (\tau )$ give a time-dependent particle definition. 
The instantaneous mode functions 
$\phi _{\mathbf{p} _{\! \perp} ,\lambda } ^{\pm \, (\tau _1)} (\tau ,\eta ,\mathbf{x}_{\! \perp}) $ 
are related with the in mode functions
$\phi _{\mathbf{p} _{\! \perp} ,\lambda } ^{\pm \, \text{in}} (\tau ,\eta ,\mathbf{x} _{\! \perp} ) $ 
as follows: 
\begin{equation}
\begin{split}
\phi _{\mathbf{p} _{\! \perp} ,\lambda } ^{+\, \text{in}} (\tau ,\eta ,\mathbf{x} _{\! \perp} ) 
 &= \int \! d^2 p_{\! \perp}^\prime d\lambda^\prime \left[ 
    \left( \phi _{\mathbf{p}_{\! \perp}^\prime ,\lambda^\prime } ^{+\, (\tau _1)} ,
    \phi _{\mathbf{p}_{\! \perp} ,\lambda } ^{+\, \text{in}} \right) _\tau
    \phi _{\mathbf{p}_{\! \perp}^\prime ,\lambda^\prime } ^{+\, (\tau _1)} (\tau ,\eta ,\mathbf{x}_{\! \perp})
    -\left( \phi _{\mathbf{p}_{\! \perp}^\prime ,\lambda^\prime } ^{-\, (\tau _1)} ,
    \phi _{\mathbf{p}_{\! \perp} ,\lambda } ^{+\, \text{in}} \right) _\tau 
    \phi _{\mathbf{p}_{\! \perp}^\prime ,\lambda^\prime } ^{-\, (\tau _1)} (\tau ,\eta ,\mathbf{x}_{\! \perp})
    \right] \\
 &= \alpha _{\mathbf{p}_{\! \perp} ,\lambda +eA_\eta (\tau_1) -eA_\eta (\tau_0) } (\tau_1 ) \,
    \phi _{\mathbf{p}_{\! \perp} ,\lambda +eA_\eta (\tau_1) -eA_\eta (\tau_0) } ^{+\, (\tau _1)} (\tau ,\eta ,\mathbf{x}_{\! \perp})
    \\ 
 &\hspace{10pt}  
    +\beta ^* _{\mathbf{p}_{\! \perp} ,\lambda +eA_\eta (\tau_1) -eA_\eta (\tau_0) } (\tau_1 ) \,
    \phi _{\mathbf{p}_{\! \perp} ,\lambda +eA_\eta (\tau_1) -eA_\eta (\tau_0) } ^{-\, (\tau _1)} (\tau ,\eta ,\mathbf{x}_{\! \perp})
\end{split} \label{in-t1-a} 
\end{equation}
\begin{equation}
\begin{split}
\phi _{\mathbf{p} _{\! \perp} ,\lambda } ^{-\, \text{in}} (\tau ,\eta ,\mathbf{x} _{\! \perp} ) 
 &= \int \! d^2 p_{\! \perp}^\prime d\lambda^\prime \left[ 
    \left( \phi _{\mathbf{p}_{\! \perp}^\prime ,\lambda^\prime } ^{+\, (\tau _1)} ,
    \phi _{\mathbf{p}_{\! \perp} ,\lambda } ^{-\, \text{in}} \right) _\tau
    \phi _{\mathbf{p}_{\! \perp}^\prime ,\lambda^\prime } ^{+\, (\tau _1)} (\tau ,\eta ,\mathbf{x}_{\! \perp})
    -\left( \phi _{\mathbf{p}_{\! \perp}^\prime ,\lambda^\prime } ^{-\, (\tau _1)} ,
    \phi _{\mathbf{p}_{\! \perp} ,\lambda } ^{-\, \text{in}} \right) _\tau
    \phi _{\mathbf{p}_{\! \perp}^\prime ,\lambda^\prime } ^{-\, (\tau _1)} (\tau ,\eta ,\mathbf{x}_{\! \perp})
    \right] \\
 &= \beta _{\mathbf{p}_{\! \perp} ,\lambda +eA_\eta (\tau_1) -eA_\eta (\tau_0) } (\tau_1 ) \, 
    \phi _{\mathbf{p}_{\! \perp} ,\lambda +eA_\eta (\tau_1) -eA_\eta (\tau_0) } ^{+\, (\tau _1)} (\tau ,\eta ,\mathbf{x}_{\! \perp})
    \\
 &\hspace{10pt}
    +\alpha ^* _{\mathbf{p}_{\! \perp} ,\lambda +eA_\eta (\tau_1) -eA_\eta (\tau_0) } (\tau_1 ) \, 
    \phi _{\mathbf{p}_{\! \perp} ,\lambda +eA_\eta (\tau_1) -eA_\eta (\tau_0) } ^{-\, (\tau _1)} 
    (\tau ,\eta ,\mathbf{x}_{\! \perp}) ,
\end{split} \label{in-t1-b}
\end{equation}
where the Bogoliubov coefficients $\alpha _{\mathbf{p} _{\! \perp} ,\lambda } (\tau ) $
and $\beta _{\mathbf{p} _{\! \perp} ,\lambda } (\tau ) $ are introduced by \footnote{
Actually, we should write $\alpha _{\mathbf{p} _{\! \perp} ,\lambda } (\tau ,\tau_1 )$ 
instead of $\alpha _{\mathbf{p} _{\! \perp} ,\lambda } (\tau_1 )$ in Eq.~\eqref{def_Bogo}. 
However, we omit one time-argument for simplicity, because the instantaneous decomposition of the field [Eq.~\eqref{expansion2}]
is physically relevant only at $\tau =\tau_1$. } 
\begin{equation}
\begin{split}
\left(\phi _{\mathbf{p} _{\! \perp} ^\prime ,\lambda ^\prime } ^{+\, (\tau_1 )} 
,\phi _{\mathbf{p} _{\! \perp} ,\lambda } ^{+\, \text{in}} \right)_\tau 
 &= \alpha _{\mathbf{p} _{\! \perp} ^\prime ,\lambda ^\prime } (\tau_1 ) 
   \delta ^2 (\mathbf{p} _{\! \perp} -\mathbf{p} _{\! \perp} ^\prime) 
   \delta (\lambda -\lambda ^\prime +eA_\eta (\tau_1 ) -eA_\eta (\tau_0) ) , \\ 
\left(\phi _{\mathbf{p} _{\! \perp} ^\prime ,\lambda ^\prime } ^{+\, (\tau_1 )} 
,\phi _{\mathbf{p} _{\! \perp} ,\lambda } ^{-\, \text{in}} \right)_\tau 
 &= \beta _{\mathbf{p} _{\! \perp} ^\prime ,\lambda ^\prime } (\tau_1 ) 
   \delta ^2 (\mathbf{p} _{\! \perp} -\mathbf{p} _{\! \perp} ^\prime) 
   \delta (\lambda -\lambda ^\prime +eA_\eta (\tau_1 ) -eA_\eta (\tau_0) ) .
\end{split} \label{def_Bogo}
\end{equation}
For the explicit forms of the Bogoliubov coefficients, see Eqs.~\eqref{Bogo_a} and \eqref{Bogo_b}. 
These coefficients satisfy the condition
$|\alpha _{\mathbf{p} _{\! \perp} ,\lambda } |^2 -|\beta _{\mathbf{p} _{\! \perp} ,\lambda } |^2 =1$. 
By the definition of $\chi _{\mathbf{p} _{\! \perp} ,\lambda} ^{\pm \, \text{in}} (\tau)$, 
$\beta _{\mathbf{p} _{\! \perp} ,\lambda } (\tau) =0$ for $\tau <\tau_0$. 

\begin{figure*}[t]
 \begin{tabular}{cc}
  \begin{minipage}{0.5\textwidth}
   \begin{center}
    \includegraphics[scale=0.5]{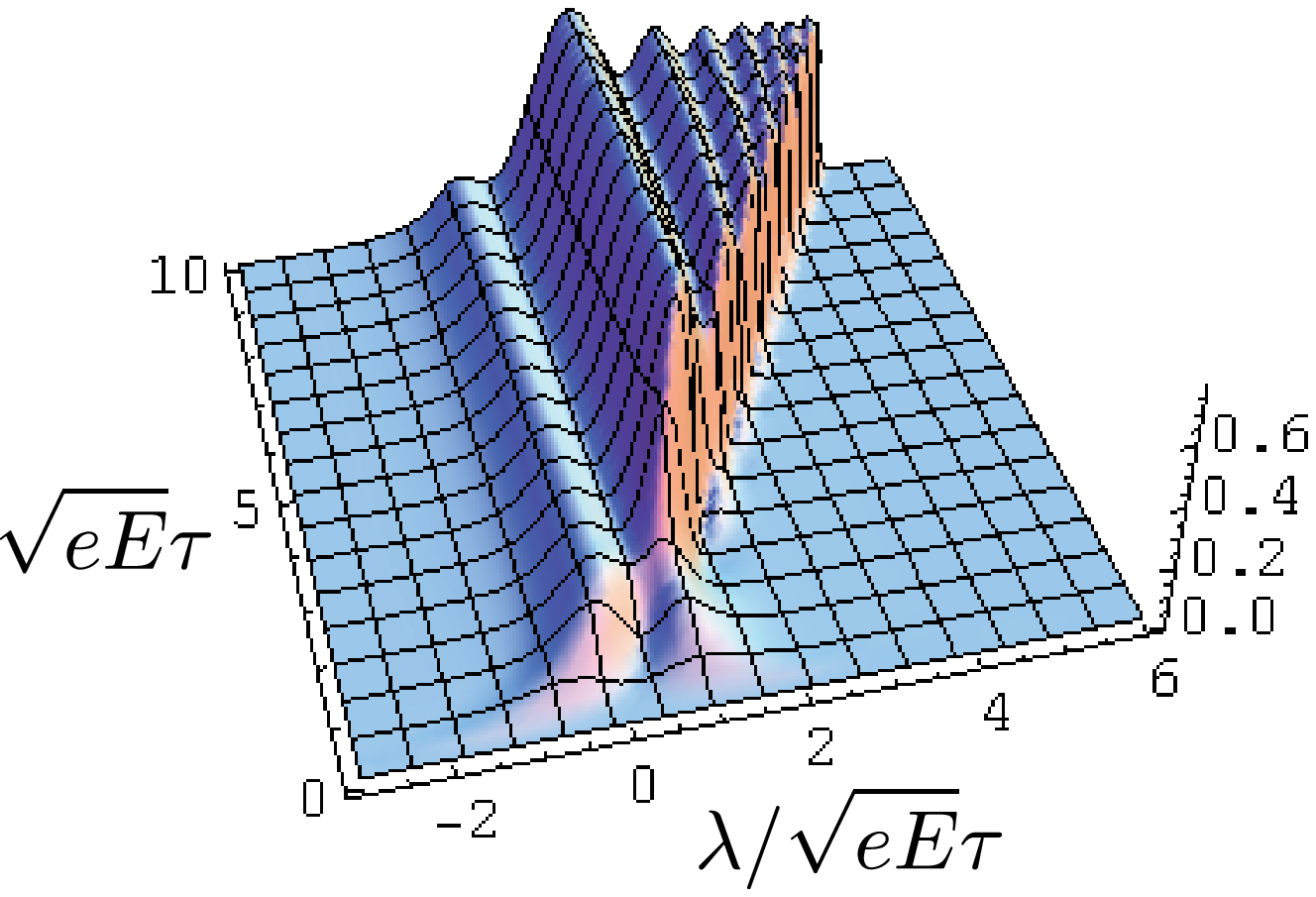}
   \end{center}
  \end{minipage} &
  \begin{minipage}{0.5\textwidth} 
   \begin{center}
    \includegraphics[scale=0.5]{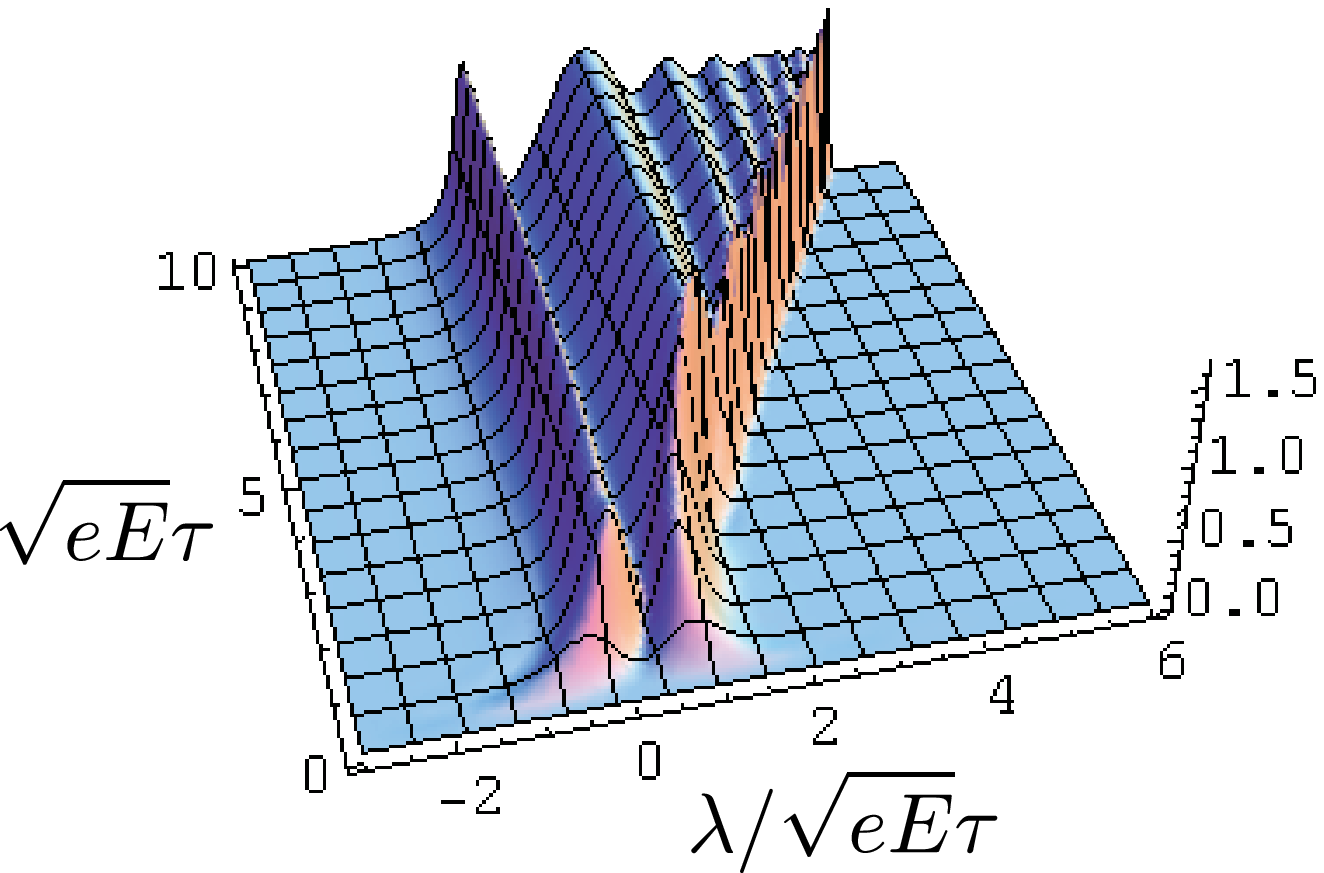}
   \end{center}
  \end{minipage} \\
  (a) $a=\frac{m_{\! \perp} ^2}{2eE} =0.1$ &
  (b) $a=\frac{m_{\! \perp} ^2}{2eE} =0.01$ 
 \end{tabular}
 \caption{(color online). Longitudinal momentum distributions with fixed transverse momentum ($\sqrt{eE} \tau _0 =0.1$) }
 \label{fig:distriL1}
\end{figure*}

Inserting Eqs.~\eqref{in-t1-a} and \eqref{in-t1-b} into Eq.~\eqref{expansion1} and comparing it with Eq.~\eqref{expansion2},
we get the relation between the particle definition associated with in-solutions and the instantaneous particle definition:
\begin{equation}
\begin{split}
\mathfrak{a}_{\mathbf{p} _{\! \perp} ,\lambda } (\tau ) 
 &= \alpha _{\mathbf{p} _{\! \perp} ,\lambda } (\tau ) 
   \mathfrak{a}_{\mathbf{p} _{\! \perp} ,\lambda -eA_\eta (\tau) +eA_\eta (\tau_0) } ^\text{in} 
   +\beta _{\mathbf{p} _{\! \perp} ,\lambda } (\tau ) 
    \mathfrak{b}_{-\mathbf{p} _{\! \perp} ,-\lambda +eA_\eta (\tau) -eA_\eta (\tau_0) } ^{\text{in} \ \dagger} , \\
\mathfrak{b}_{-\mathbf{p} _{\! \perp} ,-\lambda } ^\dagger (\tau )
 &= \alpha _{\mathbf{p} _{\! \perp} ,\lambda } ^* (\tau ) 
   \mathfrak{b}_{-\mathbf{p} _{\! \perp} ,-\lambda +eA_\eta (\tau) -eA_\eta (\tau_0) } ^{\text{in} \ \dagger} 
   +\beta _{\mathbf{p} _{\! \perp} ,\lambda } ^* (\tau ) 
    \mathfrak{a}_{\mathbf{p} _{\! \perp} ,\lambda -eA_\eta (\tau) +eA_\eta (\tau_0) } ^\text{in} .
\end{split} \label{Bogo}
\end{equation}

Because the creation and the annihilation operators are mixed by the Bogoliubov transformation \eqref{Bogo}, the vacuum expectation 
of the number operator can be nonzero:
\begin{equation}
\begin{split}
\frac{dN}{d^2 p _{\! \perp} d\lambda } &=
\langle 0,\text{in} |\mathfrak{a}_{\mathbf{p} _{\! \perp} ,\lambda } ^\dagger (\tau ) 
\mathfrak{a}_{\mathbf{p} _{\! \perp} ,\lambda } (\tau ) |0,\text{in} \rangle \\
 &= |\beta _{\mathbf{p} _{\! \perp} ,\lambda } (\tau) |^2 \frac{V_\eta }{\tau (2\pi)^3} . 
\end{split}
\end{equation}
This means that particles are created from the electric field. 
Because of the charge and the momentum conservation, antiparticles have always opposite momentum to particles:
\begin{equation}
\langle 0,\text{in} |\mathfrak{b}_{\mathbf{p} _{\! \perp} ,\lambda } ^\dagger (\tau ) 
\mathfrak{b}_{\mathbf{p} _{\! \perp} ,\lambda } (\tau ) |0,\text{in} \rangle
 = |\beta _{-\mathbf{p} _{\! \perp} ,-\lambda } (\tau) |^2 \frac{V_\eta }{\tau (2\pi)^3} .
\end{equation}

\begin{figure}
\begin{center}
\includegraphics[scale=0.6]{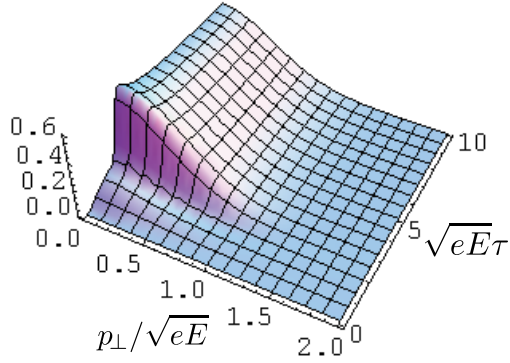} 
\end{center}
 \vskip -\lastskip \vskip -3pt
\caption{(color online). Transverse momentum distribution with fixed longitudinal momentum $\lambda /\sqrt{eE}\tau =1$ 
         ($m^2/2eE =0.1$ and $\sqrt{eE} \tau _0 =0.1$)}
\label{fig:distriT1}
\end{figure}

In Figs.~\ref{fig:distriL1},\ref{fig:distriT1}, the proper-time evolution of the distribution function
\begin{equation}
f_{\mathbf{p} _{\! \perp} ,\lambda } (\tau ) 
 = (2\pi )^3 \frac{dN}{d^2 x _{\! \perp} d\eta d^2 p _{\! \perp} d\lambda }
 = |\beta _{\mathbf{p} _{\! \perp} ,\lambda } (\tau) |^2 
\end{equation}  
is plotted. 
Figure \ref{fig:distriL1} shows the longitudinal momentum distributions with fixed transverse momentum $p_{\! \perp } =0$
for $m^2/2eE =0.1$ and 0.01. 
The longitudinal momentum distributions are plotted as a function of $\lambda /\tau $ instead of $\lambda $
because $\lambda /\tau $ denotes a momentum in the $\eta$-frame as explained in Sec.~\ref{sec:meaning}. 
Figure \ref{fig:distriT1} exhibits the transverse momentum distribution with fixed longitudinal momentum
$\lambda /\sqrt{eE} \tau =1$ for $m^2/2eE =0.1$. 
Hereafter, all figures are shown in a dimensionless unit scaled by $\sqrt{eE}$. 

These momentum distributions are very similar to those in the uniform electric fields (see Ref.~\cite{Tanji09}): 
\begin{itemize}
  \setlength{\itemsep}{0cm} %
 \item Particles are created with approximately zero longitudinal momentum. 
 \item In the transverse direction, the distributions are nearly Gaussian $\exp \left(-\frac{\pi m_{\! \perp } ^2}{eE} \right)$. 
 \item After being created, they are accelerated to the longitudinal direction 
       by the electric field according to the classical equation of motion. 
\end{itemize}
The similarity of the transverse momentum distributions is quite reasonable, 
because the fields are free in the transverse direction also in the present case. 
In contrast, the similarity of the longitudinal momentum distributions is only a superficial one.
The physical contents are quite different between the uniform electric field case and the present case,
because the physical meanings of the longitudinal momentum are distinct. 
In the uniform case, the longitudinal momentum is defined in the center of mass frame. 
On the other hand, in the present case,  the longitudinal momentum $\lambda /\tau $ denotes
the momentum observed in the $\eta$-frame, which is the frame boosted by \eqref{boost} from the center of mass frame. 
Therefore, Fig.~\ref{fig:distriL1} does not mean particles are created with zero longitudinal momentum 
in the center of mass frame, 
but it means particles are created with the scaling velocity distributions. 
That is, a particle created at the point $(t,z)$ has the velocity $z/t$ 
from the first instance when it is created. 
This result supports the assumption in Refs.~\cite{Kajantie-Matsui,Gatoff-K-M} that a source term in a kinetic equation 
contains the factor $\delta (y-\eta )$. 
In our field-theoretical treatment, however, the velocity is not exactly $v_z =z/t$ because of quantum fluctuation.  
After being created, particles undergo the acceleration by the field and their velocity distribution deviates from 
the scaling one, of which processes are expressed by $\lambda =\frac{1}{2} eE\tau ^2 +\text{const.} $ 
in a classical level.

\begin{figure}
\begin{center}
\includegraphics[scale=0.65]{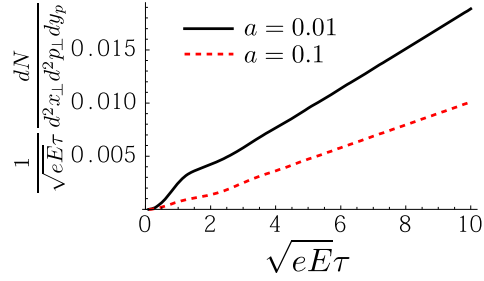} 
\end{center}
 \vskip -\lastskip \vskip -3pt
\caption{(color online). Time evolution of the particle number density $(a=\frac{m_{\! \perp} ^2}{2eE}$ and $\sqrt{eE}\tau_0 =0.1 )$}
\label{fig:number}
\end{figure}

The momentum distribution with respect to momentum defined in the Cartesian coordinates,
in other words, momentum in the center of mass frame, is calculated as follows:
\begin{equation}
\begin{split}
\frac{dN}{d^3 p} &= \langle 0,\text{in} |a_\mathbf{p} ^\dagger a_\mathbf{p} |0,\text{in} \rangle \\
 &= \frac{1}{2\pi \omega _p } \int \! d\lambda \int \! d\lambda ^\prime e^{-i(\lambda -\lambda ^\prime )y_p } 
    \langle 0,\text{in} |\mathfrak{a}_{\mathbf{p} _{\! \perp} ,\lambda } ^\dagger (\tau ) 
    \mathfrak{a}_{\mathbf{p} _{\! \perp} ,\lambda ^\prime } (\tau ) |0,\text{in} \rangle \\ 
 &= \frac{1}{2\pi \omega _p } \int \! d\lambda \ 
    f_{\mathbf{p} _{\! \perp} ,\lambda } (\tau) \frac{L^2 }{(2\pi )^2} ,
\end{split} \label{one_p}
\end{equation}
in which derivation Eq.~\eqref{relation} was used. 
Thus, the rapidity distribution
\begin{equation}
\frac{dN}{d^2 x_{\! \perp} d^2 p_{\! \perp} dy_p }
 = \omega _p \frac{1}{L^2 } \frac{dN}{d^3 p}
 = \frac{1}{(2\pi )^3 } \int \! d\lambda \ f_{\mathbf{p} _{\! \perp} ,\lambda } (\tau ) \label{one_y}
\end{equation} 
is independent of rapidity. 
Because we have assumed the perfect boost invariance, the central plateau in the rapidity distribution
extends to infinity. 
In reality, the speed of nuclei after a collision is not exactly the speed of light, 
so that the configuration of the electric field is not perfectly boost-invariant, and
there should be some cutoff in the plateau.  
Note that this independence of rapidity is a direct consequence of Eq.~\eqref{relation} and is not
affected by the explicit form of $f_{\mathbf{p} _{\! \perp} ,\lambda } (\tau )$. 
Therefore, even if the electric field has proper-time dependence or the back-reaction is taken into account, 
the rapidity distribution \eqref{one_y} is always independent of rapidity 
as long as the boost invariance of the electric field is assumed.

In Fig.~\ref{fig:number}, the time evolution of the rapidity distributions \eqref{one_y} divided by $\tau$ is plotted. 
They show linear increase at later time. 
That is, the particle number density increases quadratically in proper time. 
This is because 
(i) pair creation happens constantly some time after the switch-on of the field
and (ii) the space volume the electric field spans increases linearly in $\tau$. 

\begin{figure*}
 \begin{tabular}{cc}
  \begin{minipage}{0.5\textwidth}
   \begin{center}
    \includegraphics[scale=0.6]{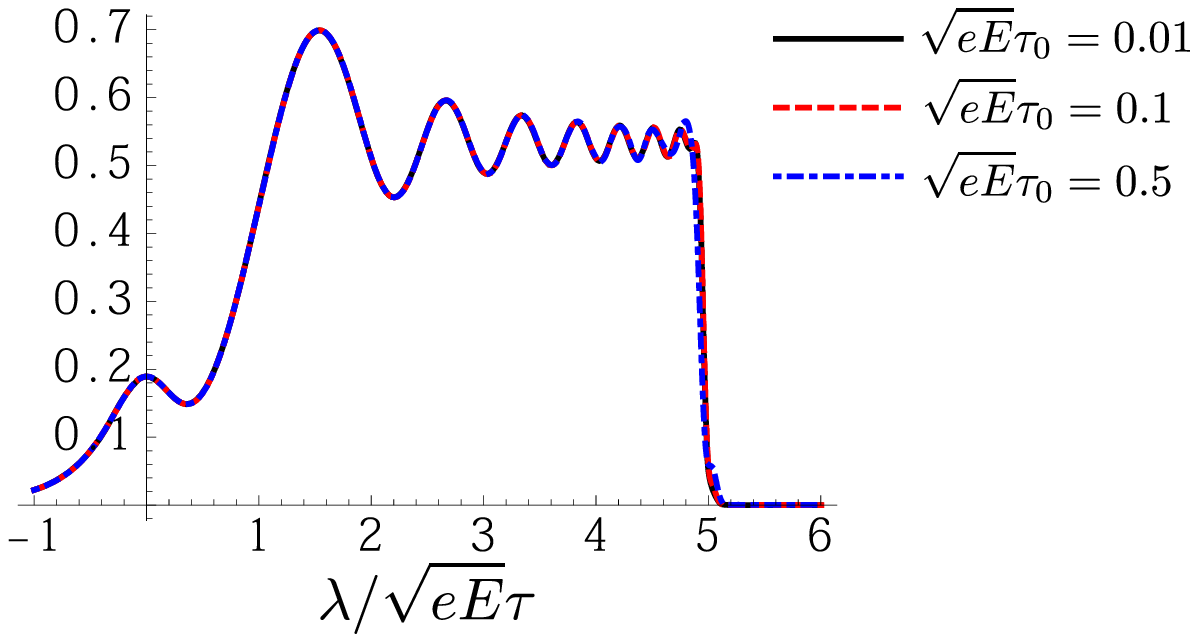}
    \\
    (a) Longitudinal momentum distribution $f_{\mathbf{p}_{\! \perp} ,\lambda} (\tau)$ at fixed time $\sqrt{eE} \tau=10$
   \end{center} 
  \end{minipage} &
  \begin{minipage}{0.5\textwidth} 
   \begin{center}
    \includegraphics[scale=0.6]{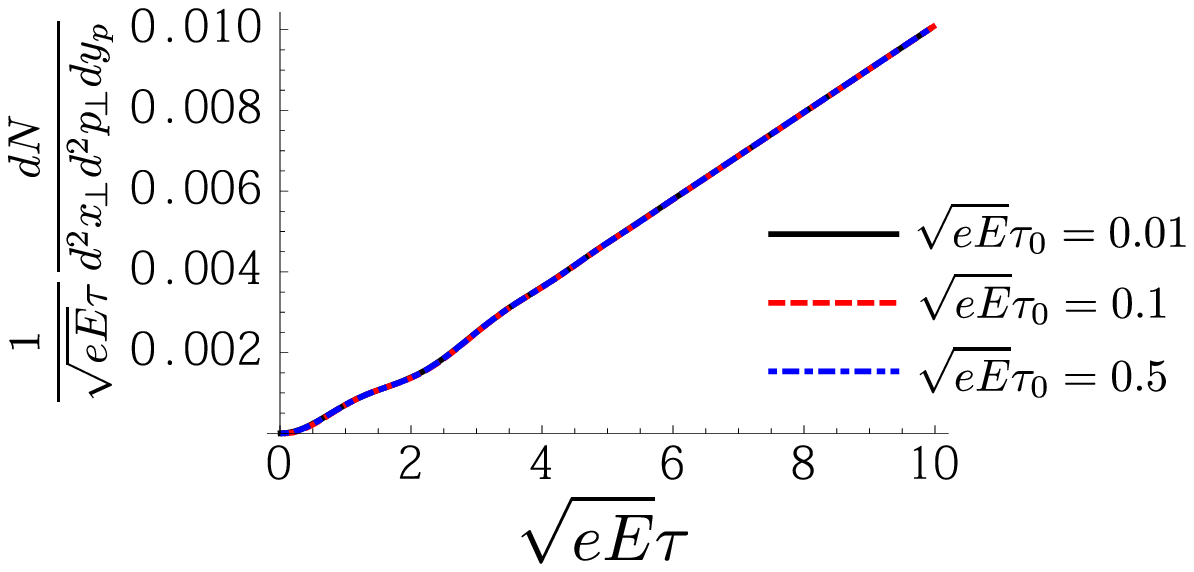}
    \\
    (b) Time evolution of the particle number density
   \end{center} 
  \end{minipage} 
 \end{tabular}
 \caption{(color online). Dependence on the time of the switch-on $\tau_0$ \ ($a=0.1$)}
 \label{fig:t-dep}
\end{figure*}

The dependence on the time of the switch-on $\tau_0$ is shown in Fig.~\ref{fig:t-dep}. 
Figure \ref{fig:t-dep}(a) represents the longitudinal momentum distribution at fixed time $\sqrt{eE} \tau=10$, 
and Fig.~\ref{fig:t-dep}(b) does the time evolution of the particle number density. 
Three lines corresponding to $\tau _0 =0.01 ,0.1$ and 0.5 are almost overlapped in both figures. 
These demonstrate insensitivity to the values of $\tau_0 $ for sufficiently small $\tau_0$ 
($\sqrt{eE} \tau_0 \lesssim 0.1$). 

\section{Two-particle correlations} \label{sec:correlation} 
Because particles are created as a coherent superposition of several momentum modes 
in the electric field which exists only inside the forward light cone,
they are correlated in the momentum space. 
In this section, we study the two-particle correlations in the momentum space
between particles created from the boost-invariant field. 

The two-particle spectrum is defined as
\begin{equation}
 \frac{dN_2}{d^3 p d^3 q} 
  = \langle 0,\text{in} |a_\mathbf{p} ^\dagger a_\mathbf{q} ^\dagger a_\mathbf{p} a_\mathbf{q} |0,\text{in} \rangle .
\end{equation}
In the case of the pair creation in a spatially uniform electric field \cite{Fukushima09}, 
the two-particle spectrum is
\begin{equation}
 \frac{dN_2}{d^3 p d^3 q} = \frac{dN}{d^3 p} \frac{dN}{d^3 q} 
  + \left\{ f_\mathbf{p} (t) \right\} ^2 \frac{V}{(2\pi)^3} \delta ^3 (\mathbf{p} -\mathbf{q} ) , \label{two_uni}
\end{equation}
where $f_\mathbf{p} (t)$ is the momentum distribution function of particles created from the field. 
The first term in the right-hand side is an uncorrelated part, 
which is a product of one-particle distribution 
$\frac{dN}{d^3 p} = \langle 0,\text{in} |a_\mathbf{p} ^\dagger a_\mathbf{p} |0,\text{in} \rangle 
 =f_\mathbf{p} (t) \frac{V}{(2\pi)^3}$, 
and the second term is a correlated part. 
Particles are correlated only if they have identical momenta.
This correlation is due to the Bose--Einstein statistics. 

Let us see how this two-particle spectrum is modified if the electric field localizes in the forward light cone. 
Using Eqs.~\eqref{relation} and \eqref{Bogo}, one can derive
\begin{equation} \label{two_p}
 \frac{dN_2}{d^3 p d^3 q} = \frac{dN}{d^3 p} \frac{dN}{d^3 q} 
   + \frac{1}{\omega _p \omega _q } 
   \left| \int \! \frac{d\lambda}{2\pi} e^{-i\lambda (y_p -y_q)} f_{\mathbf{p}_{\! \perp} ,\lambda} (\tau) \right|^2 
   \frac{L^2}{(2\pi)^2} \delta ^2 (\mathbf{p} _{\! \perp} -\mathbf{q} _{\! \perp}) , 
\end{equation}
instead of Eq.~\eqref{two_uni}. 
Converting the longitudinal momenta to rapidities gives 
\begin{equation} \label{two_y}
\frac{dN_2}{d^2 p_{\! \perp} dy_p d^2 q_{\! \perp} dy_q } 
  = \frac{dN}{d^2 p_{\! \perp} dy_p} \frac{dN}{d^2 q_{\! \perp} dy_q} 
    + \left| \int \! \frac{d\lambda}{2\pi} e^{-i\lambda (y_p -y_q)} f_{\mathbf{p}_{\! \perp} ,\lambda} (\tau) \right|^2
    \frac{L^2}{(2\pi)^2} \delta ^2 (\mathbf{p} _{\! \perp} -\mathbf{q} _{\! \perp}) . 
\end{equation} 
For later reference, we introduce the correlation function and its longitudinal part as
\begin{equation}
\begin{split}
C(y_p ,\mathbf{p} _{\! \perp} ;y_q ,\mathbf{q} _{\! \perp}) &\equiv
  \frac{\frac{dN_2}{d^2 p_{\! \perp} dy_p d^2 q_{\! \perp} dy_q } 
  -\frac{dN}{d^2 p_{\! \perp} dy_p} \frac{dN}{d^2 q_{\! \perp} dy_q} }{\frac{dN}{d^2 p_{\! \perp} dy_p} \frac{dN}{d^2 q_{\! \perp} dy_q}} \\
 &= \frac{\left| \int \! \frac{d\lambda}{2\pi} e^{-i\lambda (y_p -y_q)} f_{\mathbf{p}_{\! \perp} ,\lambda} (\tau) \right|^2}
    {\left\{ \int \! \frac{d\lambda}{2\pi} f_{\mathbf{p}_{\! \perp} ,\lambda} (\tau) \right\} ^2 }
    \frac{\delta ^2 (\mathbf{p} _{\! \perp} -\mathbf{q} _{\! \perp})}{L^2 /(2\pi )^2} ,
\end{split}
\end{equation} 
\begin{equation}
 C_L (\Delta y=y_p -y_q ,\mathbf{p} _{\! \perp} ) \equiv
  \frac{\left| \int \! \frac{d\lambda}{2\pi} e^{-i\lambda \Delta y} f_{\mathbf{p}_{\! \perp} ,\lambda} (\tau) \right|^2}
  {\left\{ \int \! \frac{d\lambda}{2\pi} f_{\mathbf{p}_{\! \perp} ,\lambda} (\tau) \right\} ^2 } . \label{C_L}
\end{equation} 
From Eq.~\eqref{two_p}
we can see that the correlation in the transverse direction is the same as that in Eq.~\eqref{two_uni}:
$\frac{L^2}{(2\pi)^2} \delta ^2 (\mathbf{p} _{\! \perp} -\mathbf{q} _{\! \perp})$. 
This correspondence is reasonable because in the transverse direction the field is free in both cases. 
This correlation is short-range in momentum space. 
What is remarkable is the correlation in the longitudinal direction. 
The correlated part no longer contains the delta function $\delta (p_z -q_z )$ 
and is given by the Fourier transform of the momentum distribution : 
$\int \! \frac{d\lambda}{2\pi} e^{-i\lambda (y_p -y_q)} f_{\mathbf{p}_{\! \perp} ,\lambda} (\tau)$. 
Therefore, particles are correlated even if their longitudinal momenta are different, 
while in Eq.~\eqref{two_uni} particles are correlated only if they have an identical momentum. 
Because the rapidity correlation is given by the Fourier transform of the momentum distribution, 
its width is approximately the inverse of the width of momentum distribution $f_{\mathbf{p}_{\! \perp} ,\lambda} (\tau)$ 
in $\lambda$ space. 
In particular, if the distribution is proportional to $\delta (\lambda )$, in other words, 
if the particles' velocity distribution is the scaling one, 
$C_L (\Delta y,\mathbf{p} _{\! \perp} ) =1$ for any $\Delta y$ and the rapidity-correlation range is infinite. 

\begin{figure*}
 \begin{tabular}{cc}
  \begin{minipage}{0.5\textwidth}
   \begin{center}
    \includegraphics[scale=0.6]{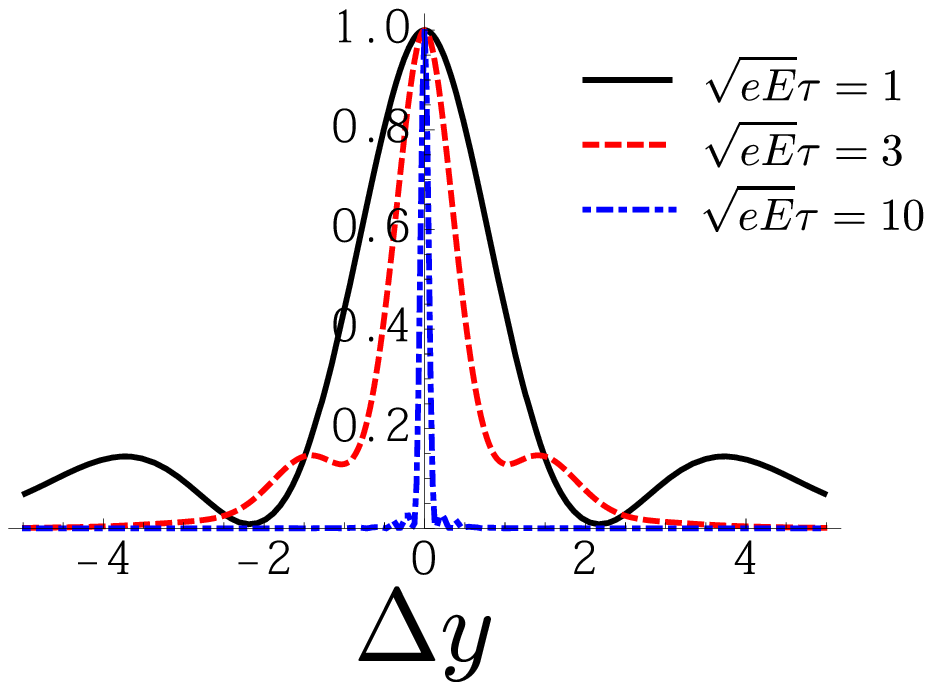}
    \vspace{5pt}
   \end{center}
  \end{minipage} &
  \begin{minipage}{0.5\textwidth} 
   \begin{center}
    \includegraphics[scale=0.6]{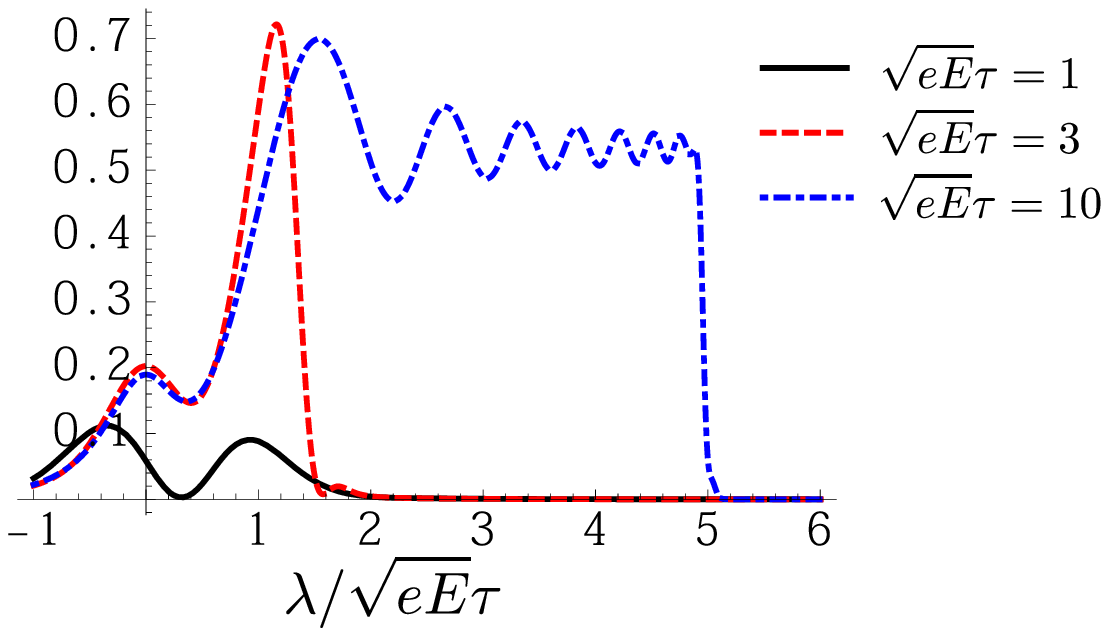}
   \end{center}
  \end{minipage} \\
  (a) the longitudinal rapidity correlations $C_L (\Delta y)$ &
  (b) the longitudinal distributions $f_{\mathbf{p}_{\! \perp} ,\lambda} (\tau)$
 \end{tabular}
 \caption{(color online). Time dependence of the correlations and the corresponding momentum distributions 
          $(m_\perp ^2/2eE =0.1$ and $\sqrt{eE} \tau _0 =0.1)$}
 \label{fig:corr-time}
\end{figure*}

\begin{figure*}
 \begin{tabular}{cc}
  \begin{minipage}{0.5\textwidth}
   \begin{center}
    \includegraphics[scale=0.6]{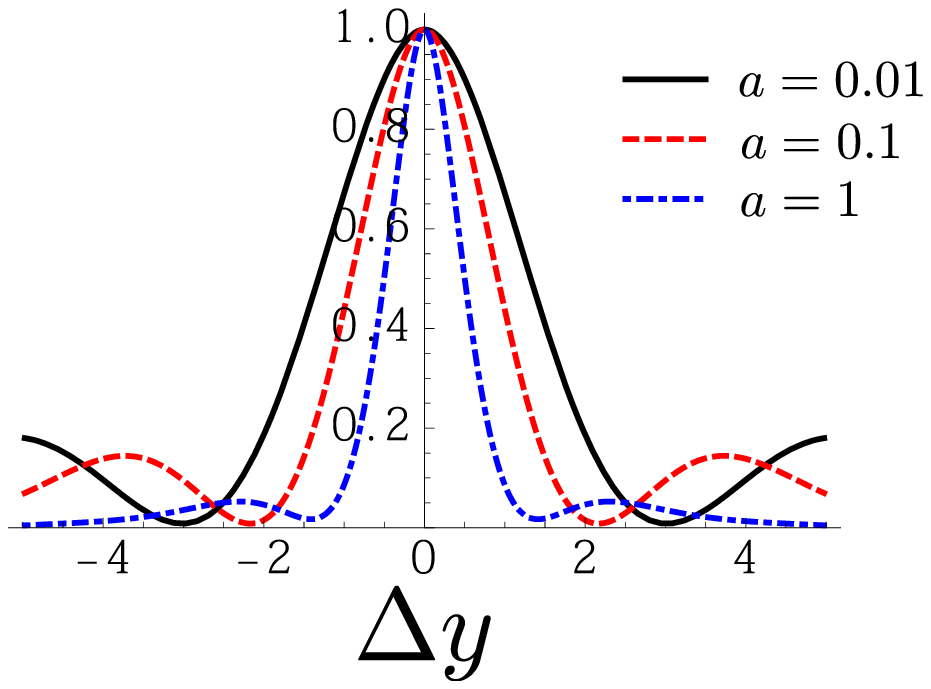}
    \vspace{5pt}
   \end{center}
  \end{minipage} &
  \begin{minipage}{0.5\textwidth} 
   \begin{center}
    \includegraphics[scale=0.6]{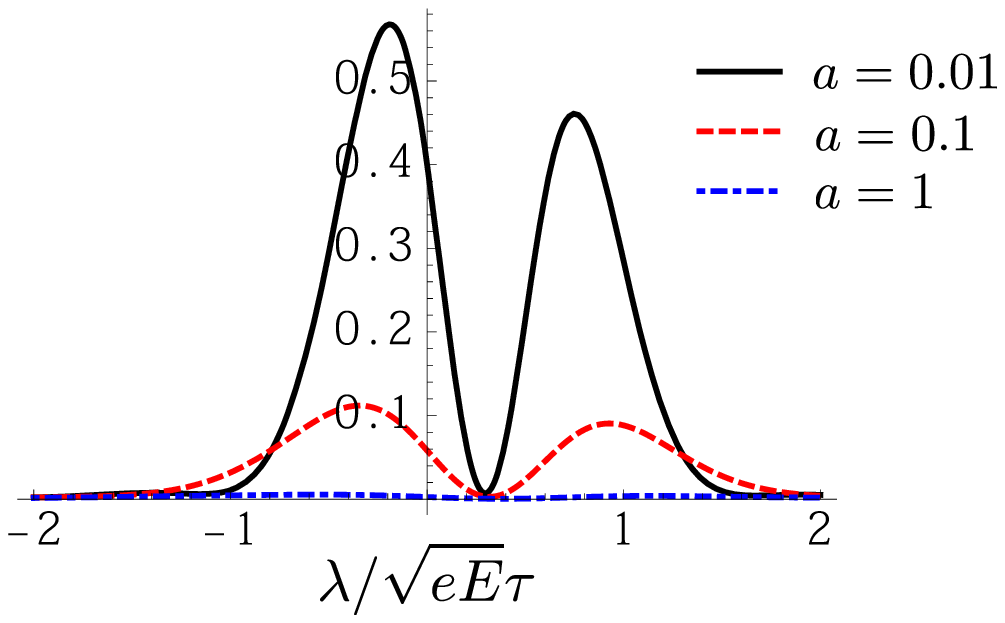}
   \end{center}
  \end{minipage} \\
  (a) the longitudinal rapidity correlations $C_L (\Delta y)$  &
  (b) the momentum distributions $f_{\mathbf{p}_{\! \perp} ,\lambda} (\tau)$ 
 \end{tabular}
 \caption{(color online). $a=\frac{m_{\! \perp}^2}{2eE}$ dependence of the correlations 
          and the corresponding momentum distributions $(\sqrt{eE} \tau =1$ and $\sqrt{eE} \tau _0 =0.1)$ }
 \label{fig:corr-a}
\end{figure*}

Figure \ref{fig:corr-time} exhibits the time evolution of (a) the longitudinal correlation $C_L$
under the constant electric field and (b) the corresponding momentum distributions $f_{\mathbf{p}_{\! \perp} ,\lambda} $, 
whose Fourier transformations give the correlation functions through Eq.~\eqref{C_L}. 
Because of the pair creation and subsequent acceleration, the width of the longitudinal momentum distribution grows in time. 
As a result, the longitudinal correlation fades away as time goes on. 
This can be understood as follows: 
Because particles which are created at the same time are strongly correlated, the correlation is long-range at first. 
As particle production continues to happen, many particles which are created at
different times begin to coexist, so that the correlations are attenuated at later times. 

In Fig.~\ref{fig:corr-a}, the $a=\frac{m_{\! \perp}^2}{2eE}$ dependence of (a) the longitudinal correlations
at fixed time $\sqrt{eE} \tau =1$ and (b) the corresponding momentum distributions are shown. 
The two peak structures seen in the momentum distributions at early time are owing to some quantum effects and
have been obtained in earlier works treating the pair creation of scalar particles \cite{Tanji09,Fukushima09}. 
The rapidity-correlation length decreases with increasing $a$. 
This is because the height of the momentum distributions depends on $a$ exponentially ($\sim e^{-2\pi a}$) \cite{Tanji09} 
and their width scaled by $\sqrt{eE}$ is nearly independent of $a$,
so that the relative width of the momentum distributions increases with increasing $a$.  
This $a$ dependence means that the heavier the transverse mass or the weaker the electric field,
the shorter the rapidity correlation. 

Before closing this section, let us note that although we have dealt with the constant electric field which is independent of $\tau$,
the formula \eqref{two_p} is valid for electric fields having arbitrary $\tau$ dependence. 
Thus, we can use Eq.~\eqref{two_p} even if the back-reaction of the particle production is taken into account,
within the approximation that interaction between produced particles through quantum gauge fields is neglected. 

\section{Summary and discussion} \label{summary}
In this paper, we have studied the pair creation of scalar particles in electric fields which span only inside
the forward light cone as a model for plasma formation in heavy-ion collisions. 
We have assumed the electric field is symmetric under the Lorentz boost along the longitudinal beam direction 
and investigated how particle production happens maintaining the boost invariance of the field. 

To give a description which holds the boost invariance, we have developed the field quantization in terms of 
the curvilinear $\tau$-$\eta$ coordinates. 
This formalism may be useful for not only the study of pair creation but also for other studies
that quantum fields in a boost-invariantly expanding system are involved. 
Although we have treated scalar fields for simplicity, generalization to Dirac fields or vector fields
is straightforward. 
A detailed formulation of Dirac fields in the $\tau$-$\eta$ coordinates has been presented in Ref.~\cite{Mihaila06}. 

We have shown that under the boost-invariant field, particles are created as a coherent superposition of
several eigenmodes of the longitudinal momentum. 
As a result, the rapidity distribution of the particles is independent of rapidity,
and the particles have the scaling velocity distribution $v_z =z/t$ from the first instance they are created. 
A significant point is that this flow of particles consists of quantum-mechanical states. 
Each particle spreads in the forward light cone 
and forms the scaling velocity distribution as a quantum state.
It is not the case that a cluster of classical particles forms this flow like in the Bjorken flow in classical hydrodynamics. 
This fact brings nontrivial rapidity correlations among the particles. 

We have calculated the two-particle correlations between particles created from the boost-invariant field,
and found that the correlation is short-range with respect to the transverse momentum, 
which originates in the Bose--Einstein correlation, and is long-range with respect to the longitudinal rapidity. 
These features may remind us of the near-side ridge phenomena observed 
in nucleus-nucleus collisions at the RHIC \cite{STAR05,PHENIX08,PHOBOS10} 
and recently in proton-proton collisions at the LHC \cite{CMS}. 
However, to make quantitative predictions, further improvements of this study, such as 
introducing a proper-time dependence of the electric field due to expansion of the system and
taking the back reaction of the pair creation into account, are necessary. 
The back reaction can be introduced by the Maxwell equation
\begin{equation}
\frac{1}{\tau} \frac{d}{d\tau} \left( \frac{1}{\tau} \frac{d}{d\tau} A_\eta (\tau ) \right) 
 = -\langle 0,\text{in} | j^\eta |0,\text{in} \rangle , \label{Maxwell}
\end{equation}
where $j^\eta $ denotes the charge current operator of the produced particles. 
In our calculation, the longitudinal correlation fades away as proper time goes on,
because under the constant field the particle production continues to happen and 
many particles which are created at different times coexist at later times. 
Hence, we can conclude that the life time of the electric field should be short
to obtain a long-range rapidity correlation. 
Furthermore, considering a tubelike structure of the electric field and effects of radial flow may be important 
for the transverse correlation \cite{Dumitru08}. 
It is also interesting to study whether the correlation can survive under a thermalization process,
which is not included in our treatment, or under the subsequent hydrodynamic evolution.

\section*{Acknowledgments}
The author thanks Professor T. Matsui for enlightening discussions and careful reading of the manuscript 
and Professor H. Fujii for helpful discussions and comments on the manuscript.  
Comments by Professor T. Hirano are greatly appreciated. 
The author also thanks Yoshiaki Onishi for his kind comments and discussions. 
The author is supported by the Japan Society for the Promotion of Science for Young Scientists.

\appendix
\section{Solutions of the Klein--Gordon equation under the boost-invariant field} \label{sec:solutions}
In this appendix, we show a set of solutions of the Klein--Gordon equation [Eq.~\eqref{KGinE1}] 
\begin{equation}
\left[ \tau ^2 \partial_\tau ^2 +\tau \partial_\tau -D_\eta ^2 
      -\tau ^2 \partial_{\! \perp} ^2 +m^2 \tau ^2 \right] 
      \phi_{\mathbf{p} _{\! \perp} ,\lambda } ^\pm (\tau ,\eta ,\mathbf{x} _{\! \perp}) =0 \label{appKG}
\end{equation} 
under the gauge $A_\eta =\frac{1}{2} E\tau ^2 $ with the initial condition such that
they are continuous with 
\begin{equation}
\phi _{\mathbf{p} _{\! \perp} ,\lambda } ^{\pm \, \text{in}} (\tau ,\eta ,\mathbf{x} _{\! \perp}) 
 = \chi _{\mathbf{p} _{\! \perp} ,\lambda } ^{\pm \, (0)} (\tau )
   \frac{1}{\sqrt{(2\pi)^3}} e^{i\mathbf{p}_{\! \perp} \cdot \mathbf{x}_{\! \perp} 
   +i(\lambda -\frac{1}{2} E\tau_0^2 )\eta} , \label{ininAp} 
\end{equation}
respectively, at $\tau =\tau_0$, where
\begin{gather}
\chi _{\mathbf{p} _{\! \perp} ,\lambda } ^{+\, (0)} (\tau )
 =\frac{\sqrt{\pi }}{2i} e^{\frac{\pi}{2} \lambda } H_{i\lambda } ^{(2)} (m_{\! \perp} \tau ) \label{+chi0} \\
\chi _{\mathbf{p} _{\! \perp} ,\lambda } ^{-\, (0)} (\tau ) 
 = \left[ \chi _{\mathbf{p} _{\! \perp} ,\lambda } ^{+\, (0)} (\tau ) \right] ^* . \label{-chi0}
\end{gather}

Under the gauge potential $A_\eta = \frac{1}{2} E\tau ^2 $, 
a set of normalized solutions of Eq.~\eqref{appKG} 
can be constructed by the confluent hypergeometric functions $U(a,b,z)$ \cite{Abramowitz} as follows:
\begin{equation}
 \tilde{\phi} _{\mathbf{p} _{\! \perp} ,\lambda } ^\pm (\tau ,\eta ,\mathbf{x} _{\! \perp})
  = \tilde{\chi} _{\mathbf{p} _{\! \perp} ,\lambda } ^\pm (\tau )
    \frac{1}{\sqrt{(2\pi)^3}} e^{i\mathbf{p}_{\! \perp} \cdot \mathbf{x}_{\! \perp}} e^{i\lambda \eta} ,
\end{equation}
where
\begin{gather}
\begin{split}
 \tilde{\chi} _{\mathbf{p} _{\! \perp} ,\lambda } ^+ (\tau)
  &= \frac{1}{\sqrt{2} } e^{-\frac{\pi }{2} (a+\lambda )} \left( \frac{eE}{2} \tau^2 \right) ^{\frac{i}{2} \lambda } 
    e^{-\frac{i}{4} eE\tau ^2 } 
    U(1/2 +ia+i\lambda ,1+i\lambda ,ieE\tau ^2 /2 ) 
\end{split} \label{chiAp} \\
 \tilde{\chi} _{\mathbf{p} _{\! \perp} ,\lambda } ^- (\tau )
  = \left[ \tilde{\chi} _{\mathbf{p} _{\! \perp} ,\lambda } ^+ (\tau ) \right] ^* \ .
\end{gather}
These solutions $\tilde{\phi}_{\mathbf{p} _{\! \perp} ,\lambda } ^\pm $ are not continuous
with Eq.~\eqref{ininAp} at $\tau =\tau_0$. 
To find solutions which are connected smoothly to Eq.~\eqref{ininAp} at $\tau =\tau_0$,
we decompose the solutions $\tilde{\phi} _{\mathbf{p} _{\! \perp} ,\lambda } ^\pm $ by
$\phi _{\mathbf{p} _{\! \perp} ,\lambda } ^{\pm \, (\tau_1 )}$, which are defined 
by Eq.~\eqref{phi^t1}: 
\begin{equation}
\begin{split}
\tilde{\phi} _{\mathbf{p} _{\! \perp} ,\lambda } ^+ (\tau ,\eta ,\mathbf{x} _{\! \perp}) 
 &= \int \! d^2 p_{\! \perp} ^\prime d\lambda ^\prime \left[
    \left( \phi ^{+\, (\tau_1 )} _{\mathbf{p} _{\! \perp}^\prime ,\lambda^\prime } ,
    \tilde{\phi} _{\mathbf{p} _{\! \perp} ,\lambda } ^+ \right) _{\! \tau}
    \phi ^{+\, (\tau_1 )} _{\mathbf{p} _{\! \perp}^\prime ,\lambda^\prime } (\tau ,\eta ,\mathbf{x} _{\! \perp}) 
    -
    \left( \phi ^{-\, (\tau_1 )} _{\mathbf{p} _{\! \perp}^\prime ,\lambda^\prime } ,
    \tilde{\phi} _{\mathbf{p} _{\! \perp} ,\lambda } ^+ \right) _{\! \tau}
    \phi ^{-\, (\tau_1 )} _{\mathbf{p} _{\! \perp}^\prime ,\lambda^\prime } (\tau ,\eta ,\mathbf{x} _{\! \perp}) 
    \right] \\
 &= A_{\mathbf{p} _{\! \perp} ,\lambda +eA_\eta (\tau_1)} (\tau,\tau_1 ) \, 
    \phi ^{+\, (\tau_1 )} _{\mathbf{p} _{\! \perp} ,\lambda +eA_\eta (\tau_1)} 
    (\tau ,\eta ,\mathbf{x} _{\! \perp}) \\
 & \hspace{12pt}
    +B_{\mathbf{p} _{\! \perp} ,\lambda +eA_\eta (\tau_1)} ^* (\tau,\tau_1 ) \, 
    \phi ^{-\, (\tau_1 )} _{\mathbf{p} _{\! \perp} ,\lambda +eA_\eta (\tau_1)} 
    (\tau ,\eta ,\mathbf{x} _{\! \perp}) 
\end{split} \label{+tildeexp}
\end{equation}
\begin{equation}
\begin{split}
\tilde{\phi} _{\mathbf{p} _{\! \perp} ,\lambda } ^- (\tau ,\eta ,\mathbf{x} _{\! \perp}) 
 &= \int \! d^2 p_{\! \perp} ^\prime d\lambda ^\prime \left[
    \left( \phi ^{+\, (\tau_1 )} _{\mathbf{p} _{\! \perp}^\prime ,\lambda^\prime } ,
    \tilde{\phi} _{\mathbf{p} _{\! \perp} ,\lambda } ^- \right) _{\! \tau}
    \phi ^{+\, (\tau_1 )} _{\mathbf{p} _{\! \perp}^\prime ,\lambda^\prime } (\tau ,\eta ,\mathbf{x} _{\! \perp}) 
    -
    \left( \phi ^{-\, (\tau_1 )} _{\mathbf{p} _{\! \perp}^\prime ,\lambda^\prime } ,
    \tilde{\phi} _{\mathbf{p} _{\! \perp} ,\lambda } ^- \right) _{\! \tau}
    \phi ^{-\, (\tau_1 )} _{\mathbf{p} _{\! \perp}^\prime ,\lambda^\prime } (\tau ,\eta ,\mathbf{x} _{\! \perp}) 
    \right] \\
 &= B_{\mathbf{p} _{\! \perp} ,\lambda +eA_\eta (\tau_1)} (\tau,\tau_1 ) \, 
    \phi ^{+\, (\tau_1 )} _{\mathbf{p} _{\! \perp} ,\lambda +eA_\eta (\tau_1)} 
    (\tau ,\eta ,\mathbf{x} _{\! \perp}) \\
 & \hspace{12pt}
    +A_{\mathbf{p} _{\! \perp} ,\lambda +eA_\eta (\tau_1)} ^* (\tau,\tau_1 ) \, 
    \phi ^{-\, (\tau_1 )} _{\mathbf{p} _{\! \perp} ,\lambda +eA_\eta (\tau_1)} 
    (\tau ,\eta ,\mathbf{x} _{\! \perp}) ,
\end{split} \label{-tildeexp}
\end{equation}
where the Bogoliubov coefficients $A_{\mathbf{p} _{\! \perp} ,\lambda} (\tau,\tau_1 ) $
and $B_{\mathbf{p} _{\! \perp} ,\lambda} (\tau,\tau_1 ) $ are defined, respectively, by
\begin{gather}
A_{\mathbf{p} _{\! \perp} ,\lambda} (\tau,\tau_1 ) 
 = i\tau \left\{ \left[ \chi _{\mathbf{p} _{\! \perp} ,\lambda } ^{+\, (0)} (\tau ) \right] ^*
    \overleftrightarrow{\frac{d}{d\tau}} 
    \tilde{\chi} _{\mathbf{p} _{\! \perp} ,\lambda -eA_\eta (\tau_1)} ^+ (\tau ) \right\} \label{BogoA} \\
B_{\mathbf{p} _{\! \perp} ,\lambda} (\tau,\tau_1 ) 
 = i\tau \left\{ \left[ \chi _{\mathbf{p} _{\! \perp} ,\lambda } ^{+\, (0)} (\tau ) \right] ^*
    \overleftrightarrow{\frac{d}{d\tau}} 
    \tilde{\chi} _{\mathbf{p} _{\! \perp} ,\lambda -eA_\eta (\tau_1)} ^- (\tau ) \right\} . \label{BogoB}
\end{gather}
Substituting Eqs.~\eqref{+chi0}, \eqref{-chi0} and \eqref{chiAp} into these equations, 
one can derive the explicit forms of the Bogoliubov coefficients. 
Albeit Eqs.~\eqref{+tildeexp} and \eqref{-tildeexp} are mathematically valid for any positive $\tau$ and $\tau_1$, 
the expansion by the mode functions 
$\phi ^{\pm \, (\tau_1 )} _{\mathbf{p} _{\! \perp} ,\lambda} (\tau ,\eta ,\mathbf{x} _{\! \perp})$
is physically meaningful only at $\tau =\tau_1$. 
Therefore, it is sufficient to know the Bogoliubov coefficients having the same time argument $\tau =\tau _1$: 
$A_{\mathbf{p} _{\! \perp} ,\lambda} (\tau ) \equiv A_{\mathbf{p} _{\! \perp} ,\lambda} (\tau,\tau ) $ and 
$B_{\mathbf{p} _{\! \perp} ,\lambda} (\tau ) \equiv B_{\mathbf{p} _{\! \perp} ,\lambda} (\tau,\tau ) $. 
After completing the differentiation in Eqs.~\eqref{BogoA} and \eqref{BogoB}, we can set $\tau =\tau_1$
and obtain the following equations: 
\begin{gather}
\begin{split}
A_{\mathbf{p} _{\! \perp} ,\lambda} (\tau ) 
 &= \frac{\sqrt{2\pi}}{4} e^{-\frac{\pi}{2} (a-\frac{1}{2} eE\tau ^2 )} 
    \left( \frac{1}{2} eE\tau ^2 \right) ^{\frac{i}{2} \left( \lambda -\frac{1}{2} eE\tau ^2 \right) }
    e^{-\frac{i}{4} eE\tau ^2} \\
 &\times \left\{ 
    (1+2ia) H_{-i\lambda } ^{(1)} (m_{\! \perp} \tau ) 
    U(\frac{1}{2} +ia+i(\lambda -\frac{1}{2} eE\tau ^2 ),i(\lambda -\frac{1}{2} eE\tau ^2 ),\frac{i}{2} eE\tau ^2 ) 
    \right. \\
 & \hspace{20pt} \left. 
    -m_{\! \perp} \tau H_{1-i\lambda } ^{(1)} (m_{\! \perp} \tau ) 
    U(\frac{1}{2} +ia+i(\lambda -\frac{1}{2} eE\tau ^2 ),1+i(\lambda -\frac{1}{2} eE\tau ^2 ),\frac{i}{2} eE\tau ^2 )
    \right\}
\end{split} \\
\begin{split}
B_{\mathbf{p} _{\! \perp} ,\lambda} ^* (\tau ) 
 &= \frac{\sqrt{2\pi}}{4} e^{-\frac{\pi}{2} (a-\frac{1}{2} eE\tau ^2 )} 
    \left( \frac{1}{2} eE\tau^2 \right) ^{\frac{i}{2} \left( \lambda -\frac{1}{2} eE\tau ^2 \right) }
    e^{-\frac{i}{4} eE\tau ^2} \\
 &\times \left\{ 
    (1+2ia) H_{i\lambda } ^{(2)} (m_{\! \perp} \tau ) 
    U(\frac{1}{2} +ia+i(\lambda -\frac{1}{2} eE\tau ^2 ),i(\lambda -\frac{1}{2} eE\tau ^2 ),\frac{i}{2} eE\tau ^2 ) 
    \right. \\
 & \hspace{20pt} \left. 
    +m_{\! \perp} \tau H_{1+i\lambda } ^{(2)} (m_{\! \perp} \tau ) 
    U(\frac{1}{2} +ia+i(\lambda -\frac{1}{2} eE\tau ^2 ),1+i(\lambda -\frac{1}{2} eE\tau ^2 ),\frac{i}{2} eE\tau ^2 )
    \right\} .
\end{split}
\end{gather}

Taking the relations \eqref{+tildeexp} and \eqref{-tildeexp}, 
we can construct the mode functions $\phi _{\mathbf{p} _{\! \perp} ,\lambda } ^\pm $ 
which are continuous with Eq.~\eqref{ininAp} 
(which is equivalent to $\phi _{\mathbf{p} _{\! \perp} ,\lambda } ^{\pm \, (\tau_0 )} $)
at $\tau =\tau_0 $ as follows:
\begin{gather}
\phi _{\mathbf{p} _{\! \perp} ,\lambda } ^+ (\tau ,\eta ,\mathbf{x} _{\! \perp}) 
 = A_{\mathbf{p} _{\! \perp} ,\lambda} ^* (\tau_0 ) \,
   \tilde{\phi} _{\mathbf{p} _{\! \perp} ,\lambda -eA(\tau_0 )} ^+ (\tau ,\eta ,\mathbf{x} _{\! \perp}) 
   -B_{\mathbf{p} _{\! \perp} ,\lambda} ^* (\tau_0 ) \,
   \tilde{\phi} _{\mathbf{p} _{\! \perp} ,\lambda -eA(\tau_0 )} ^- (\tau ,\eta ,\mathbf{x} _{\! \perp}) \\
\phi _{\mathbf{p} _{\! \perp} ,\lambda } ^- (\tau ,\eta ,\mathbf{x} _{\! \perp}) 
 = A_{\mathbf{p} _{\! \perp} ,\lambda} (\tau_0 ) \, 
   \tilde{\phi} _{\mathbf{p} _{\! \perp} ,\lambda -eA(\tau_0 )} ^- (\tau ,\eta ,\mathbf{x} _{\! \perp}) 
   -B_{\mathbf{p} _{\! \perp} ,\lambda} (\tau_0 ) \, 
   \tilde{\phi} _{\mathbf{p} _{\! \perp} ,\lambda -eA(\tau_0 )} ^+ (\tau ,\eta ,\mathbf{x} _{\! \perp}) .
\end{gather}
Further using Eqs.~\eqref{+tildeexp} and \eqref{-tildeexp}, we can expand the mode functions 
$\phi _{\mathbf{p} _{\! \perp} ,\lambda } ^\pm$ by the instantaneous free solutions at $\tau =\tau_1$:
\begin{align}
\phi _{\mathbf{p} _{\! \perp} ,\lambda } ^+ (\tau ,\eta ,\mathbf{x} _{\! \perp}) 
 &= \left\{ A_{\mathbf{p} _{\! \perp} ,\lambda} ^* (\tau_0 ) 
    A_{\mathbf{p} _{\! \perp} ,\lambda +eA_\eta (\tau_1) -eA_\eta (\tau_0)} (\tau_1 )
    -B_{\mathbf{p} _{\! \perp} ,\lambda} ^* (\tau_0 ) 
    B_{\mathbf{p} _{\! \perp} ,\lambda +eA_\eta (\tau_1) -eA_\eta (\tau_0)} (\tau_1 ) \right\} \notag \\
 &\hspace{20pt} \times
    \phi _{\mathbf{p} _{\! \perp} ,\lambda +eA_\eta (\tau_1) -eA(\tau_0 )} ^{+\, (\tau_1)}
    (\tau ,\eta ,\mathbf{x} _{\! \perp}) \notag \\
 &\hspace{12pt}
    +\left\{ A_{\mathbf{p} _{\! \perp} ,\lambda} ^* (\tau_0 ) 
    B_{\mathbf{p} _{\! \perp} ,\lambda +eA_\eta (\tau_1) -eA_\eta (\tau_0)} ^* (\tau_1 )
    -B_{\mathbf{p} _{\! \perp} ,\lambda} ^* (\tau_0 ) 
    A_{\mathbf{p} _{\! \perp} ,\lambda +eA_\eta (\tau_1) -eA_\eta (\tau_0)} ^* (\tau_1 ) \right\} \notag \\
 &\hspace{20pt} \times
    \phi _{\mathbf{p} _{\! \perp} ,\lambda +eA_\eta (\tau_1) -eA(\tau_0 )} ^{-\, (\tau_1)}
    (\tau ,\eta ,\mathbf{x} _{\! \perp}) \notag \\
 &= \alpha _{\mathbf{p} _{\! \perp} ,\lambda +eA_\eta (\tau_1) -eA_\eta (\tau_0)} (\tau_1 )
    \phi _{\mathbf{p} _{\! \perp} ,\lambda +eA_\eta (\tau_1) -eA(\tau_0 )} ^{+\, (\tau_1)}
    (\tau ,\eta ,\mathbf{x} _{\! \perp}) \notag \\
 &\hspace{12pt}
    +\beta_{\mathbf{p} _{\! \perp} ,\lambda +eA_\eta (\tau_1) -eA_\eta (\tau_0)} ^* (\tau_1 )
    \phi _{\mathbf{p} _{\! \perp} ,\lambda +eA_\eta (\tau_1) -eA(\tau_0 )} ^{-\, (\tau_1)}
    (\tau ,\eta ,\mathbf{x} _{\! \perp}) \\
\phi _{\mathbf{p} _{\! \perp} ,\lambda } ^- (\tau ,\eta ,\mathbf{x} _{\! \perp}) 
 &= \alpha _{\mathbf{p} _{\! \perp} ,\lambda +eA_\eta (\tau_1) -eA_\eta (\tau_0)} ^* (\tau_1 )
    \phi _{\mathbf{p} _{\! \perp} ,\lambda +eA_\eta (\tau_1) -eA(\tau_0 )} ^{-\, (\tau_1)}
    (\tau ,\eta ,\mathbf{x} _{\! \perp}) \notag \\
 &\hspace{12pt}
    +\beta_{\mathbf{p} _{\! \perp} ,\lambda +eA_\eta (\tau_1) -eA_\eta (\tau_0)} (\tau_1 )
    \phi _{\mathbf{p} _{\! \perp} ,\lambda +eA_\eta (\tau_1) -eA(\tau_0 )} ^{+\, (\tau_1)}
    (\tau ,\eta ,\mathbf{x} _{\! \perp}) ,
\end{align}
where the Bogoliubov coefficients
\begin{gather}
\alpha _{\mathbf{p} _{\! \perp} ,\lambda} (\tau)
 = A_{\mathbf{p} _{\! \perp} ,\lambda -eA_\eta (\tau) +eA_\eta (\tau_0)} ^* (\tau_0 ) 
   A_{\mathbf{p} _{\! \perp} ,\lambda} (\tau )
   -B_{\mathbf{p} _{\! \perp} ,\lambda -eA_\eta (\tau) +eA_\eta (\tau_0)} ^* (\tau_0 ) 
   B_{\mathbf{p} _{\! \perp} ,\lambda} (\tau ) \label{Bogo_a} \\
\beta _{\mathbf{p} _{\! \perp} ,\lambda} (\tau)
 = A_{\mathbf{p} _{\! \perp} ,\lambda -eA_\eta (\tau) +eA_\eta (\tau_0)} (\tau_0 ) 
   B_{\mathbf{p} _{\! \perp} ,\lambda} (\tau )
   -B_{\mathbf{p} _{\! \perp} ,\lambda -eA_\eta (\tau) +eA_\eta (\tau_0)} (\tau_0 ) 
   A_{\mathbf{p} _{\! \perp} ,\lambda} (\tau ) \label{Bogo_b}
\end{gather}
have been introduced.

\end{document}